\documentclass[acmsmall, nonacm=true]{acmart}
\usepackage{graphicx}
\usepackage{subcaption}
\usepackage{rotating}
\usepackage{lipsum, afterpage}
\usepackage{footnote}
\usepackage{lscape}
\usepackage{longtable}
\usepackage{array, makecell}
\usepackage{siunitx, mhchem}
\usepackage{tikz}
\usepackage{multirow}
\usepackage{adjustbox}
\usepackage{pdflscape}
\setcitestyle{square}

\AtBeginDocument{%
  \providecommand\BibTeX{{%
    \normalfont B\kern-0.5em{\scshape i\kern-0.25em b}\kern-0.8em\TeX}}}

\setcopyright{acmcopyright}
\acmJournal{CSUR}

\begin{document}

\title{Placement of Microservices-based IoT Applications in Fog Computing: A Taxonomy and Future Directions}

\author{Samodha Pallewatta}
\email{ppallewatta@student.unimelb.edu.au}
\author{Vassilis Kostakos}
\email{vassilis.kostakos@unimelb.edu.au}
\author{Rajkumar Buyya}
\email{rbuyya@unimelb.edu.au}
\affiliation{%
  \institution{The Cloud Computing and Distributed Systems (CLOUDS) Laboratory, School of Computing and Information Systems, University of Melbourne}
  \country{Australia}
}

\renewcommand{\shortauthors}{S. Pallewatta, V. Kostakos, R. Buyya}

\begin{abstract}
The Fog computing paradigm utilises distributed, heterogeneous and resource-constrained devices at the edge of the network for efficient deployment of latency-critical and bandwidth-hungry IoT application services. Moreover, MicroService Architecture (MSA) is increasingly adopted to keep up with the rapid development and deployment needs of fast-evolving IoT applications. Due to the fine-grained modularity of the microservices and their independently deployable and scalable nature, MSA exhibits great potential in harnessing Fog and Cloud resources, thus giving rise to novel paradigms like Osmotic computing. The loosely coupled nature of the microservices, aided by the container orchestrators and service mesh technologies, enables the dynamic composition of distributed and scalable microservices to achieve diverse performance requirements of the IoT applications using distributed Fog resources. To this end, efficient placement of microservice plays a vital role, and scalable placement algorithms are required to utilise the said characteristics of the MSA while overcoming novel challenges introduced by the architecture. Thus, we present a comprehensive taxonomy of recent literature on microservices-based IoT applications placement within Fog computing environments. Furthermore, we organise multiple taxonomies to capture the main aspects of the placement problem, analyse and classify related works, identify research gaps within each category, and discuss future research directions. 
\end{abstract}




\maketitle

\section{Introduction}
The Internet of Things (IoT) paradigm is gaining immense popularity due to its significance in technical, social and economic aspects \cite{bhardwaj2016review}. IoT transforms everyday objects and infrastructure into intelligent entities that can interact with each other without human intervention, which has resulted in its expansion to a wide range of services, including healthcare, transportation, industrialisation, agriculture, etc. IoT generates enormous quantities of dynamic data composed of various data types to be processed, analysed and stored. Cloud computing was initially identified as a viable solution for hosting such IoT services, giving rise to cloud-centric IoT \cite{gubbi2013internet}. However, due to the exponential increase in connected devices, raw data transmission towards centralised Cloud data centres increases network congestion and latency, thus reducing the feasibility of cloud-centric IoT analytics. As a solution, distributed computing paradigms like Fog computing, Edge computing, and Mobile Edge Computing (MEC) are introduced, bringing data processing and storage closer to the end-user, hence supporting latency-critical and bandwidth-consuming IoT services. 

The Fog computing paradigm, first introduced by Cisco in 2012, uses compute, storage, and networking services residing in the intermediate layer between IoT devices and the Cloud (i.e., smart routers and switches, micro datacentres, small-cell base stations, nano servers, Raspberry Pi devices, cloudlets, etc.) \cite{bonomi2012fog}. Compared to Fog computing, Edge computing considers only the resources located exclusively at the edge of the network (i.e., mobile phones, access points, etc.) \cite{goudarzi2022scheduling}, whereas MEC focuses on the edge of the mobile network by improving the capabilities of the base stations located within the Radio Access Network (RAN) of the 5G/6G networks \cite{costa2022orchestration}. Thus, Edge computing and MEC are limited to the resources in the outer layer of the hierarchical Fog layer \cite{jamil2022resource}, while Fog computing expands further to include resources in multiple hierarchical levels extending towards Cloud and the federation among those resources to support IoT application workloads (although some research works use them interchangeably \cite{damsgaard2022approximation, carrion2022kubernetes, salaht2020overview}). However, all three paradigms share common characteristics, such as proximity to the network edge, low-latency support, location awareness, geo-distribution and integration with Cloud to overcome resource limitations.

Meanwhile, MicroService Architecture (MSA) emerged as a cloud-native application architecture style, enabling the development and deployment of highly reliable and scalable software systems that can undergo frequent updates and deployments \cite{waseem2020systematic,salah2016evolution}. This paved the way for the convergence of Fog computing, IoT and microservices, thus resulting in the introduction of novel paradigms such as Osmotic Computing that focus on dynamic placement and deployment of microservices across federated Fog-Cloud environments. Microservices-based IoT applications are gaining tremendous momentum due to their potential to improve the performance of IoT services deployed within distributed computing environments \cite{naha2018fog}. Compared to previous application architectures such as monolithic architecture and Service Oriented Architecture (SOA) realised through web services, the true potential of MSA as a cloud-native application architecture lies in its loosely coupled nature, which enables containerised deployment, dynamic composition, and load-balancing across federated multi-fog and multi-cloud environments with the support of other cloud-native technologies like container orchestrators (i.e., Kubernetes, Docker Swarm) and service mesh technologies (i.e., Istio, Linkerd). Microservices can be independently and dynamically deployed and horizontally scaled across hybrid environments while maintaining seamless connectivity among interacting microservices under dynamic conditions. Thus, within geo-distributed and heterogeneous Fog environments, the placement of microservices-based applications remains one of the most critical and challenging areas. The placement algorithms benefit from awareness of the MSA-related characteristics so that they can adequately utilise the strengths of the application architecture while overcoming its challenges. To this end, we aim to understand the features and challenges related to the development of placement policies for microservices-based IoT applications within Fog environments and analyse recent literature to summarise the current status of the research and propose possible future research directions.

\subsection{Microservices-based IoT Applications and Fog Computing}

\begin{figure}[!ht]
    \includegraphics[width=\linewidth]{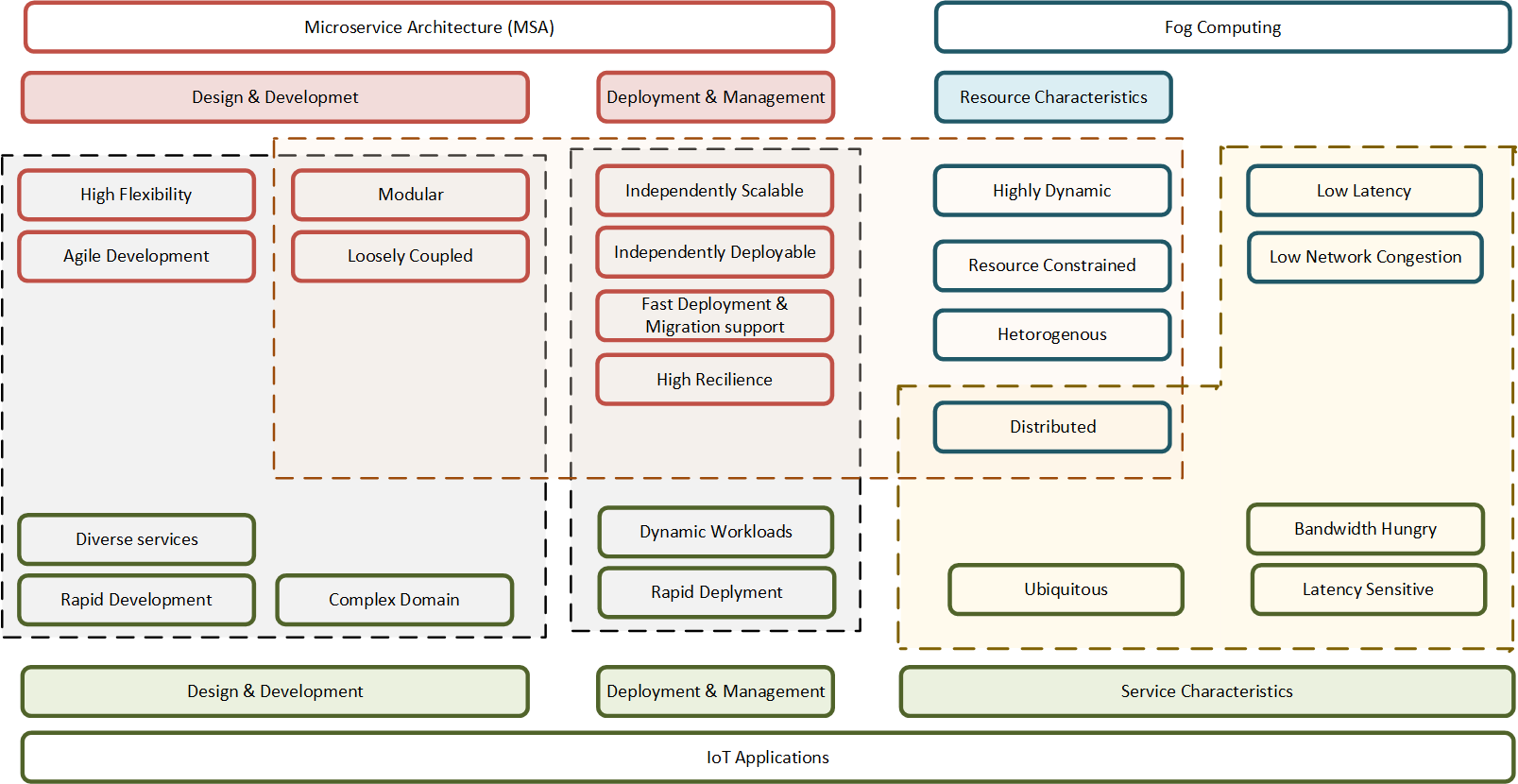}
    \caption{IoT Applications, Microservice Architecture and Fog computing}
    \label{fig:characteristic}
    \vspace*{-2.0mm}
\end{figure}

This section highlights how the main characteristics of MSA, IoT applications and Fog computing paradigm fit perfectly together, giving rise to microservices-based IoT applications for Fog environments. Fig. \ref{fig:characteristic} lists these characteristics and groups them to show their relationships. 

Characteristics of IoT Applications can be categorised under three main groups; Design and Development, Deployment and Management, and Service Characteristics. Rapid design, development needs and interoperability, and service reusability caused by the IoT ecosystem's complexity can be satisfied by adopting a software architecture that supports higher flexibility to change, reduced time to market, and collaboration among multiple development teams. MSA can enable these requirements with the fine-grained modular design and loosely coupled nature which perfectly conforms with agile design and development principles. From a deployment and maintenance perspective, IoT applications must maintain rapid deployment cycles with minimum service disruptions and support the dynamic workload while maintaining service availability and resilience. These requirements can be successfully satisfied through the use of MSA \cite{al2018enhancing, santana2018microservices, butzin2016microservices}. MSA decomposes large and complicated applications into independently deployable and scalable modules that can be conveniently packaged using container technologies such as Docker and rktlet. Lightweight containers with considerably lower startup times enable quick deployment and migration, which result in higher service availability and reliability under dynamic conditions. The decomposition of the application into small independent units allows only the updated or newly developed microservices to be re-deployed and just the performance-affected microservices to be scaled within each deployment cycle. Moreover, this enables a proper balance between horizontal and vertical scalability where microservices that are harder to scale horizontally (i.e., services that use relational databases etc.) can be vertically scaled while the rest can be horizontally scaled \cite{luksa2017kubernetes}. Thus, MSA meets the scalability, maintainability, extensibility, and interoperability requirements of large and complex IoT software systems \cite{razzaq2020systematic}. As a result, MSA is increasingly adopted for IoT application development in many areas such as smart cities, smart healthcare, IIoT \cite{thanh2021ioht, de2019microservices, bugshan2021privacy}.

Radio access network (RAN) technologies are improving rapidly to support higher bandwidth (i.e., 5g, 6g, etc.) and lower latency values. However, core network capabilities are still limited, thus making cloud-centric IoT infeasible to support IoT workloads. Sending a large amount of data towards the centralised Cloud through the core network would result in higher network congestion and larger latency values, lowering the QoS satisfaction of the latency-critical IoT services. Fog computing emerges as the solution to these challenges enabling the data analytics closer to the end devices that generates the data, thus reducing the amount of data sent towards the Cloud. Federation of Fog environments with Cloud data centres allows IoT applications to utilise the best of both paradigms. This enables latency-critical, bandwidth-hungry services to be placed closer to the end-user while the rest can use Cloud resources, thus meeting heterogeneous QoS requirements of the IoT application services. Moreover, the distributed Fog architecture creates location awareness, thus providing ubiquitous access to IoT services. With Fog resources becoming richer in resource availability and telecommunication providers supporting Fog implementation, Fog computing is rising in popularity as a viable solution for hosting IoT applications.

In contrast to Cloud data centres, Fog environments consist of distributed, heterogeneous, resource-constrained devices. Thus, deploying a monolithic application onto such devices is less feasible due to their resource demand. The scalability of such applications is also limited by the said characteristic of Fog devices. Fine-grained microservices match such environments better due to their modular, loosely coupled nature, which makes the resource requirements of each microservice small enough to be satisfied by the distributed resources. Independent deployability and scalability of such microservices enable them to adjust to dynamic conditions (i.e., device failures, mobility, workload changes, etc.) while utilising limited resources. Moreover, these characteristics support dynamic and fast migration and composition of microservices across distributed resources, thus improving deployment flexibility. Thus, MSA demonstrates the potential to utilise Fog environments and achieve a balance between federated Fog and Cloud usage to improve application performance and meet QoS requirements. 

\subsection{Microservices-based IoT Application Placement in Fog Computing Environments}

\subsubsection{Problem Definition and Challenges} \label{challenges}
Microservices-based IoT Application Placement within Fog computing environments falls under Fog Application Placement Problem (FAPP) addressed in works such as \cite{guerrero2019evaluation, skarlat2017towards, brogi2020place, salaht2020overview}. 
FAPP addresses application deployment and maintenance within Fog environments, where services with agreed Service Level Agreements (SLA) are deployed onto federated Fog and Cloud resources for the shared use by the application users. To this end, FAPP considers factors such as horizontal and vertical scalability, request load balancing, ubiquitous access, location awareness, and fault tolerance to produce application deployments that meet the required performance of the application services \cite{guerrero2019evaluation, brogi2020place}.

Based on the definition of FAPP, we define "microservices-based IoT application placement in Fog environments" as follows:

Let $A$ be a microservices-based IoT application where $A$ consists of a set of independently deployable and scalable microservices ($M_A$) and a set of services provided to the application users ($S_A$). We use the term "service" to denote end-user-requested business functionalities, which can be either an atomic service (consisting of only a single microservice) or composite services (composed of multiple microservices). As the microservices are granular with well-defined business boundaries, they ($m \in M_A$) can communicate using lightweight protocols to create composite services ($s \in S_A$) requested by the end-users. Placement of such applications includes mapping these microservices to distributed Fog and Cloud resources such that the requirements (i.e., resource requirements of the microservices, QoS requirements of the services, etc.) are ensured while maintaining seamless connectivity across distributed microservices to create composite services. Fig. \ref{fig:FSPP}  provides a motivation scenario to elaborate this further. The event flow of the Fog application placement process shown in Fig. \ref{fig:FSPP} includes distributed control engines receiving application placement requests from application providers and processing them for placement across Fog device clusters and Cloud data centres using efficient and scalable placement algorithms. Control engines can process them sequentially or as a batch based on the defined placement policy. Placement policy uses knowledge about the application model, QoS requirements, and microservice composition-related features to produce a resultant placement for the microservices.

To emphasise the novel aspects of microservices-based application placement compared to other application architectures, we identify the main challenges related to solving the placement problem as follows:


\begin{figure}[h]
    \includegraphics[width=\linewidth]{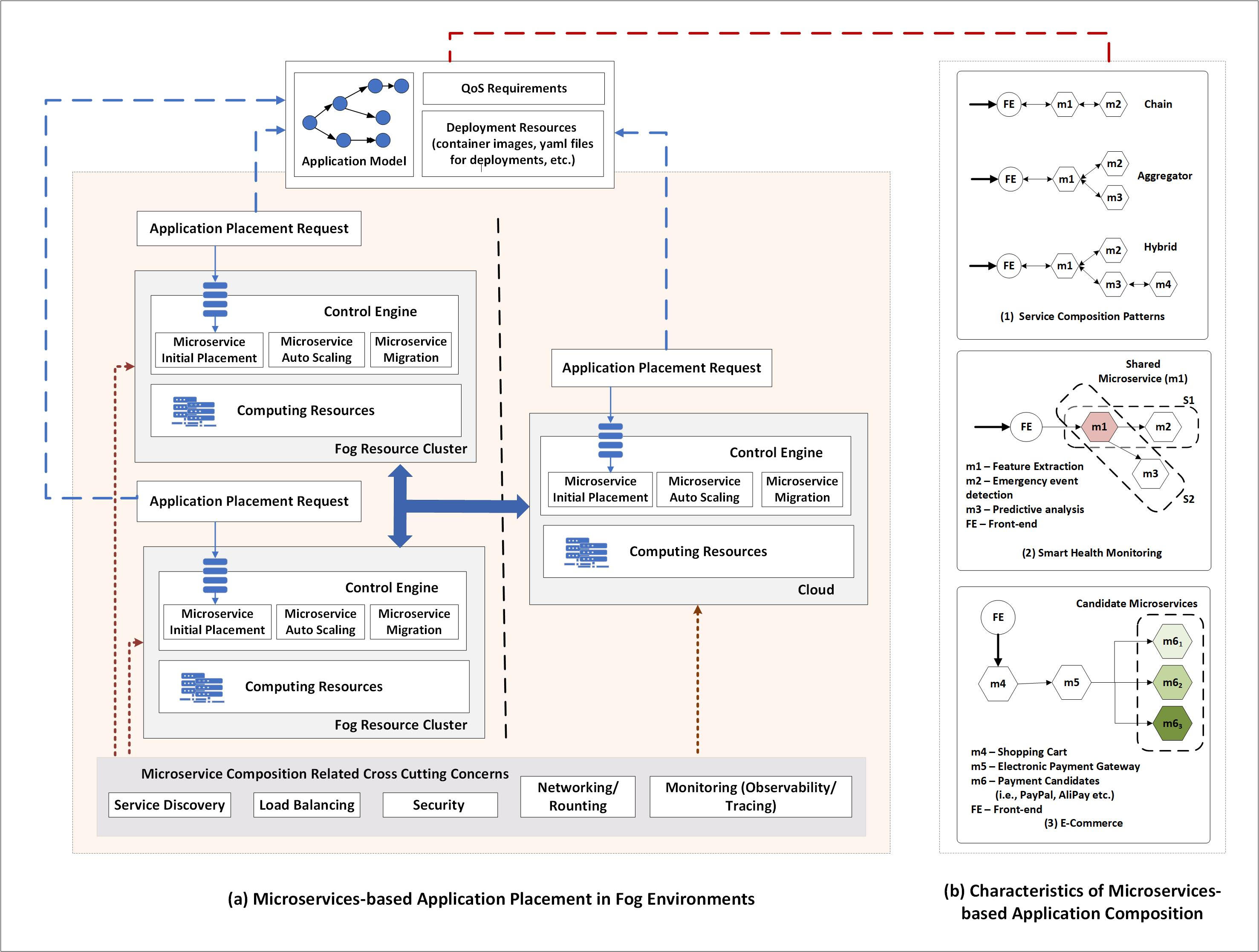}
    \caption{Microservices-based Fog Application Placement Problem}
    \label{fig:FSPP}
\end{figure}


1) \textit{\textbf{Challenges related to accurate modelling of MSA}}: The precise formulation of the placement problem depends on accurate modelling of the application architecture, such that all essential characteristics of the architecture are properly abstracted and captured to support the placement of an extensive range of IoT applications developed following the said architecture. When considering MSA, the fine-grained nature of the microservices and their complex interaction patterns set MSA apart from other application architectures and add novel challenges to the placement problem formulation. 
    
We explain these characteristics through a couple of case studies (see Fig.\ref{fig:FSPP}.b). As depicted in Fig.\ref{fig:FSPP}.b2, granular and inter-operable microservices enable applications to be developed as a collection of composite services with heterogeneous QoS requirements. Here, the Smart healthcare application for patient monitoring has two services: emergency notification service for detecting abnormalities in real-time (service S1) and latency-tolerant service for AI-based predictive analysis to give early health warnings (service S2) \cite{pallewatta2022qos}. Moreover, such granular decomposition of an application into microservices following the separation of concern principle results in heterogeneous microservices regarding resource requirements (i.e., m1 and m2 with CPU requirements and m3 with GPU requirements). Furthermore, being granular and loosely coupled components, microservices create complex interaction patterns such as shared microservices among composite services (i.e., m1 is part of both S1 and S2), candidature microservices creating conditional dataflows based on the
user demand, and third-party microservices (i.e., m6 in Fig.\ref{fig:FSPP}.b3 depicting a payment gateway with multiple optional microservices for request routing) which add more constraints on latency, security, scalability, etc. during placement. Depending on the interactions among microservices, the dataflows and possible data paths can vary too. Thus, unlike in the previous works that solved the placement and scheduling of workflows with directional acyclic data flows \cite{rodriguez2017taxonomy}, MSA can have composite services with bi-direction and cyclic dataflows where one microservice invokes other microservices, receives responses from them, and finally aggregate and further process the responses before returning the results of the composite service (see Fig.\ref{fig:FSPP}.b1) \cite{islam2022optimal, pallewatta2022qos}.

Placement policies can use the knowledge of such characteristics to utilise the distributed resources across multiple Fog clusters and Cloud data centres to overcome the challenge of limited Fog resources and meet heterogeneous resource requirements and QoS requirements while ensuring seamless connectivity among interacting microservices (i.e., placing m1 and m2 in the Fog layer closer to the network edge to meet stringent latency requirements \cite{pallewatta2019microservices} while placing m3 in the Cloud to meet the resource requirements, determining the number of horizontally scaled microservice instances to be placed for m62, m62, m63 based on their demand \cite{zhao2020distributed}, calculation of latencies depending on different dataflows within composite services \cite{pallewatta2022qos}, etc.)

Thus, to utilise MSA's full potential and overcome related challenges, accurate modelling of the applications is of paramount importance. This includes the correct depiction of microservice granularity (i.e., number of microservices, microservice heterogeneity and their invocation patterns), service composition (i.e., number of microservices per service and their dataflow patterns), and application composition (i.e., number of heterogeneous services per application and use of shared/candidature/3rd party microservices).

\textit{\textbf{Challenges related to developing microservices placement policy}}: Developing a novel placement policy includes problem formulation, which contains optimisation metrics, objectives, constraints, etc., and, based on that creating an efficient algorithm to reach an optimum placement. Compared to other application models, the loosely coupled nature and the resultant complex interaction patterns, along with higher dynamism, affects different aspects of the problem formulation, such as QoS granularity (i.e., latency requirements can be defined per composite service or among interacting microservices), QoS-awareness (i.e., handling competing QoS requirements among multiple composite services with shared microservices in a QoS-aware manner as in Fig.\ref{fig:FSPP}.b2), incorporating the horizontally and vertically scaled placement of microservices across federated Fog, and Cloud environments and load balancing requests among dynamically composed distributed instances \cite{deng2020optimal, zhao2020distributed, pallewatta2022qos} to utilise limited resources better. This provides opportunities to create efficient redundant placements, optimum request routing-based placements, locality-aware placements, resource contention-aware batch placements,  etc., to meet the expected performance of the services. 

While these aspects increase the complexity of the placement process, their incorporation in placement policies contributes to optimum and balanced utilisation of distributed, heterogeneous and resource-constrained Fog devices and resource-rich, centralised Cloud resources to improve the performance of heterogeneous IoT application services. 

\textit{\textbf{Challenges related to microservice composition}}:  The composition of granular microservices to create composite services becomes a critical challenge affecting the application's performance. The main challenges related to this include cross-cutting functions such as service discovery, load balancing, monitoring, networking etc., that come with distributed placement, scalability, elasticity, migration, redundant deployment and failures of microservices. Furthermore, as a cloud-native application architecture that can be extended to the distributed Fog layer, these composition-related functions need to support higher flexibility and dynamism across distributed environments, which separates microservice composition from previous web services-oriented SOA. Thus, container orchestration or choreography frameworks and service mesh technologies are introduced to handle the dynamic changes and maintain interconnections among microservices minimising adverse effects on service performance. 

Moreover, knowledge about their capabilities is essential in developing placement policies and, in turn, is vital in creating evaluation platforms (simulators and real-world test beds) to evaluate the performance of the developed policies. To this end, factors such as load-balancing policies, overheads due to dynamic service discovery, ability to have service discovery and load-balancing across multiple computing environments needs to be incorporated into placement problem formulation \cite{herrera2021optimal, huang2020ant, xu2020service} to achieve efficient placements. Moreover, the evaluation of placement policies should be carried out with the accurate implementation of these cross-cutting concerns to properly evaluate the placement policies.

\textit{\textbf{Challenges related to performance evaluation}}: Accurate evaluation of placement policies depends on the used Fog framework/platform and the workloads. Due to the lack of commercial Fog service providers providing Fog platforms such as IaaS, PaaS or SaaS, the evaluation of the novel placement approaches is mainly handled by simulations or small-scale Fog computing frameworks developed by researchers. Furthermore, with microservices architecture, simulators and frameworks need to be extended further with container orchestration/choreography support, MSA-related cross-cutting function support through service mesh technologies, distributed monitoring, dynamic placement policy integration, etc. Moreover, the workloads used for evaluations should adequately capture the complex characteristics of the MSA and large-scale IoT applications to achieve accurate evaluations.

\begin{figure}[h!]
    \includegraphics[width=0.6\linewidth]{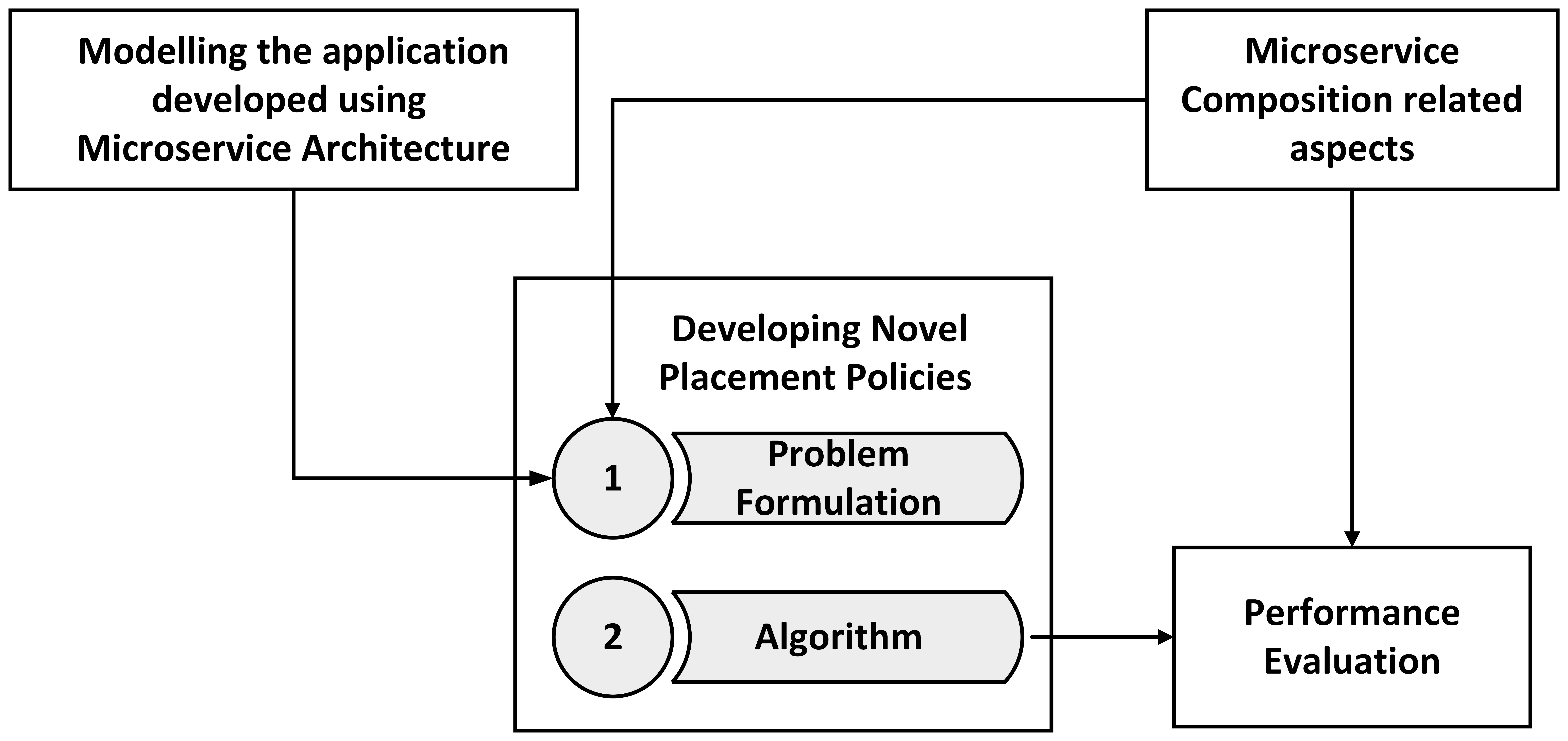}
    \caption{Main challenges related to designing novel Fog placement policies for microservice-based IoT applications and their relationships.}
    \label{fig:relationships}
    \vspace*{-3.5mm}
\end{figure} 

Fig. \ref{fig:relationships} depicts the relationships among these challenges to emphasise the importance of their collective consideration to produce efficient placements for microservices-based IoT applications.

\subsubsection{Motivation for research}

\begin{figure}[h!]
    \includegraphics[width=0.6\linewidth]{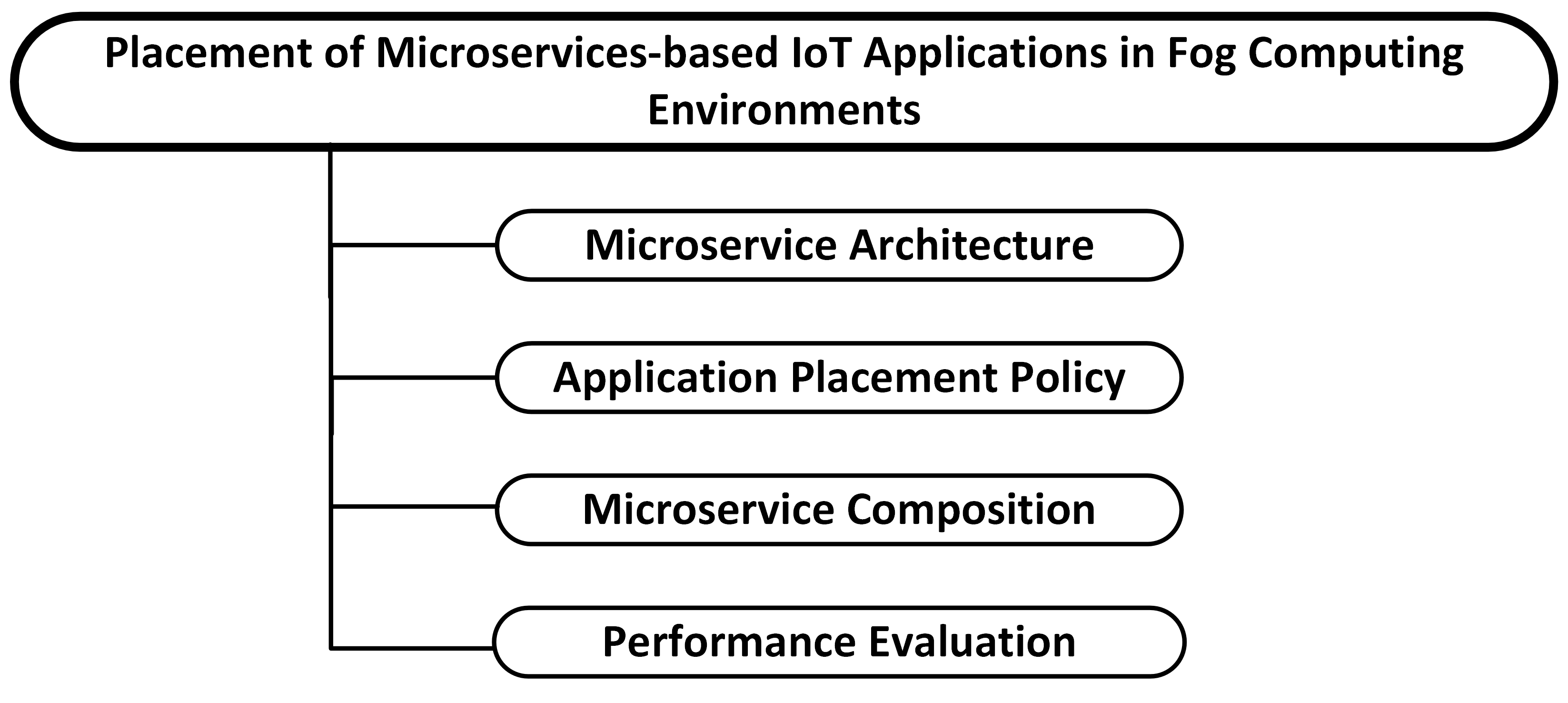}
    \caption{Taxonomy for Microservices-based IoT Applications Placement}
    \label{fig:mainTaxonomy}
    \vspace*{-3.5mm}
\end{figure} 

MSA is becoming exceedingly popular for the design and development of large-scale IoT applications, and the features of the architecture (loosely coupled nature, fine granularity, extensibility, cohesiveness, scalability, etc.) demonstrate immense potential to improve the performance of the applications through their efficient placement within federated multi-Fog multi-Cloud environments. As a cloud-native application architecture, MSA supports dynamic composition of loosely coupled microservice that can be scaled up/down across geo-distributed Fog environments and distributed across Fog resources and Cloud data centres to meet the throughput demand of the applications in a QoS-aware manner. Thus, designing efficient and scalable algorithms for the "Placement of Microservices-based IoT Applications in Fog Computing Environments" is a vital research area. As explained in the section above, to develop efficient placement algorithms, it is paramount to incorporate the characteristics, features, and challenges of the MSA into the placement problem formulation and also create evaluation platforms accordingly. Even though the research on microservices-based application placement is still in its early stages, due to the rapid adaptation of microservices in IoT application development, it is essential to analyse existing works that focus on microservices placement in Fog environments to create a comprehensive picture of the current status of research and what research gaps remain to be addressed in the future research. 

To this end, we create a taxonomy based on the characteristics and challenges discussed in previous sections (see Fig. \ref{fig:mainTaxonomy}). In the following sections, we discuss each aspect, propose separate low-level taxonomies for them, analyse existing literature under each taxonomy and identify areas of improvement. To create the taxonomy, we have carefully selected 40 research papers published in 2018-2022, with 75\% of the selected papers from 2020-2022. All the selected papers comprehensively
explore above discussed aspects related to the placement of microservices-based IoT applications
in Fog computing environments. This would provide a holistic view of multiple aspects related to problem formulation of microservices-based application placement, including; how existing works model MSA-related characteristics, how QoS parameters and constraints are modelled with respect to application architecture, what objectives are handled by the placement algorithms and how microservice characteristics facilitate them, details on microservice composition related cross-cutting functions and how they improve the dynamic nature of the microservices thus enabling better placement policies, and finally the status of evaluation platforms and workloads that can be used to evaluate placement policies accurately. Hence, this survey aims to act as a knowledge base for future research to comprehensively address the challenges of the microservices-based IoT applications placement in Fog computing environments while reaping the benefits of the architecture. The main contributions of our work are as follows:
\begin{itemize}
    \item We present a comprehensive background on MSA, IoT applications, Fog computing, and their integration to identify unique characteristics and challenges that differentiate "Microservices-based Application Placement in Fog environments" from other application models.
    \item We review recent research on microservices-based application placement in Fog environments and analyse them under multiple aspects related to the placement problem, including modelling microservices-based applications, creating application placement policies, microservice composition, and performance evaluation and propose taxonomies for each of the above perspectives. Our taxonomies capture MSA-related features and challenges in application placement.
    \item We identify research gaps in microservices-based application placement in Fog environments with reference to each proposed taxonomy.
    \item We identify and propose future research directions that researchers can use to improve the Fog application placement further.
\end{itemize}

\subsection{Paper Organization}
The rest of the paper is organised as follows. Section \ref{relatedworks} presents a qualitative analysis of existing surveys and compares them with our work. The proceeding sections introduce and discuss detailed taxonomies under each aspect identified in our high-level taxonomy: Section \ref{appModelling} introduces the taxonomy on modelling IoT applications developed according to MSA and analyses existing works, Section \ref{tax2} presents the taxonomy for placement policy design for microservices-based applications and discusses the current works accordingly, Section \ref{tax3} introduces the taxonomy for microservice composition, and Section \ref{tax4} presents the taxonomy for performance evaluation. Based on the gaps identified in each section, Section \ref{futureD} provides future research directions. Finally, Section \ref{summary} summarises this survey.

\section{Related Surveys}\label{relatedworks}

We analyse related surveys belonging to three main areas of research; surveys on Edge and Fog computing, Osmotic computing, and MSA, and conduct a qualitative comparison of the aspects covered by each of these surveys to define our contribution. Table \ref{table:relatedWork} presents a comparison of the features covered in the surveys.

Existing surveys on Edge and Fog computing cover a wide range of research areas, such as resource management, application placement, and application/task scheduling. Surveys on Fog resource management mainly focus on the architecture and characteristics of underlying Fog infrastructure (i.e., virtualisation, tenancy, etc.) and algorithms used in administrative operations, including device discovery, monitoring, performance benchmarking, load distribution, auto-scaling, etc. Fog application placement-related surveys comprehensively analyse the broad range of research work that maps applications to Fog resources to meet their non-functional requirements. In contrast, surveys on Fog application/task scheduling discuss distributing and sequencing ephemeral tasks (i.e., independent tasks or dependent tasks of the application arranged as workflows) for execution within Fog resources. These surveys also discuss works related to the computation-offloading problem in Edge and Fog computing. Thus, scheduling problem considers tasks or workflows with ephemeral life cycles that are deployed for use by a particular user, whereas Fog application placement considers deploying and managing applications with perpetual life cycles with application services accessed by a large number of users. 

Hong et al. \cite{hong2019resource} focus on resource management and, thus, study and classify the architecture of Edge and Fog platforms and algorithms designed for resource management. Their discussions on application placement are limited to analysing how static and dynamic characteristics of the resources are incorporated into placement algorithms for resource allocation. Hence, they do not focus on application architecture-related characteristics and identifying their effect on efficient placement. Jamil et al. \cite{jamil2022resource} propose a taxonomy of optimisation metrics and algorithms used in resource allocation and task scheduling in Fog computing. However, they do not characterise existing works based on the application models or architectures and fail to analyse optimisation characteristics with respect to them. Brogi et al. \cite{brogi2020place}, Salaht et al. \cite{salaht2020overview}, Mahmud et al. \cite{mahmud2020application}, and Islam et al. \cite{islam2022optimal} discuss Fog Application Placement Problem (FAPP). In Brogi et al. \cite{brogi2020place}, the main focus is on analysing placement algorithms based on the methodologies of solving the placement problem, their constraints, and optimisation metrics. Their discussion on application models is limited to modelling application module dependencies as constraints of the placement problem formulation. Salaht et al. \cite{salaht2020overview}, and Mahmud et al. \cite{mahmud2020application} provide a high-level taxonomy on covering the breadth of Fog application placement. Thus, their classifications of the application models are limited to high-level taxonomies that categorise existing works under monolithic architecture and distributed architectures such as modular and microservices. Islam et al. \cite{islam2022optimal} also provide a high-level taxonomy that classifies applications into monoliths and distributed architecture and limits the discussion on MSA to providing future research directions for microservices-based application placement. Thus, the above works do not capture MSA-related characteristics and challenges related to solving the placement problem of microservices-based IoT applications.

Goudarzi et al. \cite{goudarzi2022scheduling} also focus on providing a high-level taxonomy to give a comprehensive and broad overview of IoT application scheduling in Fog environments where the authors discuss general aspects related to scheduling, such as application structure, environmental architecture, optimisation characteristics, decision engine characteristics, and performance evaluation of algorithms. As their survey focuses on providing a high-level taxonomy of each aspect, the taxonomy of the application architecture is limited to a high-level categorisation of applications (monolithic, independent, loosely coupled, etc.). This work introduces microservices architecture as an example of loosely-coupled application structure but does not carry out further discussion on the characteristics of the architecture. An in-depth analysis of MSA is out of scope for this survey, microservice composition-related cross-cutting functionalities (i.e., load-balancing, service-discovery, networking, distributed monitoring and tracing, etc.), tools enabling dynamic composition (i.e., container orchestrators, service meshes, etc.), incorporation of MSA related features into problem formulation and algorithm evaluation (i.e., Fog computing platforms that support dynamic microservice composition, microservice workloads) are not discussed in their work. Moreover, as the main focus is on scheduling problem, FAPP-related aspects such as throughput-aware horizontal scaling, redundant placements, load-balancing, location-awareness to support distributed users, etc., are not discussed in their work.


MSA-related surveys mainly focus on development, operational phase concerns \cite{garriga2017towards, joseph2019straddling} and challenges related to adaptation of MSA for application development \cite{razzaq2020systematic}. Joseph et al. \cite{joseph2019straddling} provide a broad taxonomy using research work that captures development aspects related to MSA, such as modelling, architectural patterns, maintenance, testing and quality assurance, along with operational aspects, including placement, migration, service discovery and load balancing. Although \cite{joseph2019straddling} discusses Osmotic computing and Edge/Fog computing as distributed computing paradigms employing MSA, a categorisation of existing works within these paradigms is not performed. \cite{garriga2017towards} also provides a taxonomy covering all aspects of the microservices lifecycle. Taxonomy provided in \cite{garriga2017towards} is helpful for application developers adapting MSA to design and implement their applications, whereas our survey focuses on large-scale placement of multiple diverse microservices-based applications within distributed Fog environments. Thus, in contrast to \cite{garriga2017towards}, we explore how to incorporate microservice characteristics into the Fog application placement problem to generate optimum placements. Razzaq et al. \cite{razzaq2020systematic} study existing research to identify MSA-related software architectural styles, patterns, models, and reference architectures adapted by IoT systems. However, the placement aspects of such applications are out of the scope of their study. Thus, microservices-related surveys mainly focus on the design, development and maintenance aspects of MSA in general without focusing on challenges related to their placement and deployment within distributed computing paradigms such as Edge and Fog computing. 

Osmotic computing-related surveys focus on detailing the concepts and features of the Osmotic computing paradigm along with challenges and future directions \cite{villari2016osmotic, kaur2020osmotic}. Androcec et al. \cite{androvcec2019systematic} and Neha et al. \cite{neha2022systematic} conduct a systematic review to capture the current status of Osmotic computing by analysing applications that follow the Osmotic computing principles, Osmotic computing-related topics addressed and their level of maturity. While these works highlight the concepts and potential of Osmotic computing, they do not analyse existing works based on placement-related aspects, including features and challenges related to modelling, placement, service composition, and evaluation with respect to MSA. 

Table \ref{table:relatedWork} compares existing surveys with our work by presenting a summary of the key aspects covered. Developing efficient placement approaches for microservices-based applications requires modelling applications to capture MSA-related characteristics, placement policy creation by incorporating the application model-specific features and microservice composition-related characteristics to the placement problem formulation, and proper evaluation of the generated policies using Fog platforms that support microservice composition and microservices-based workloads. Based on the qualitative comparison, existing surveys fail to provide a thorough taxonomy to capture the above. In this work, we propose taxonomies for modelling applications based on MSA, placement policy creation, microservice composition, and performance evaluation and discuss their relationships. Moreover, we identify research gaps with respect to each aspect and propose future research directions.

 \begin{table*}[h]
	\caption{Summary of existing surveys}
	\label{table:relatedWork}
	\resizebox{\linewidth}{!}
	{\renewcommand{\arraystretch}{1.2} \begin{tabular}[htbp]{ |c|c|c|c|c|c|c|c|c|c|c|c|c|c|c|}
			\hline
			\multicolumn{1}{|c}{\textbf{}}
			&\multicolumn{5}{|c}{\textbf{ Research Areas }}
			&\multicolumn{3}{|c}{\textbf{Discussion on Application Model}} 
			&\multicolumn{4}{|c}{\textbf{MSA specific discussions}} 
			&\multicolumn{2}{|c|}{\textbf{Placement Evaluation }}
			
			\\
                \cline{2-13}
               \multicolumn{1}{|c}{\textbf{}}
			&\multicolumn{1}{|c}{\textbf{}}
			&\multicolumn{3}{|c}{\textbf{Edge/Fog Computing}} 
			&\multicolumn{1}{|c}{\textbf{}} 
			&\multicolumn{1}{|c}{\textbf{}} 
                &\multicolumn{2}{|c}{\textbf{Depth}} 
			&\multicolumn{1}{|c}{\textbf{}} 
                &\multicolumn{1}{|c}{\textbf{}} 
                &\multicolumn{1}{|c}{\textbf{Relates}} 
                &\multicolumn{1}{|c}{\textbf{Relates microservice}} 
			&\multicolumn{2}{|c|}{\textbf{}}
			
			\\
			\cline{3-5} \cline{8-9} \cline{14-15}
			\textbf{Work} &\textbf{MSA} & \textbf{Resource} & \textbf{Application} & \textbf{Application/} & \textbf{Osmotic} & \textbf{Available} & \textbf{High Level} & \textbf{Architecture}  & \textbf{Modelling}  & \textbf{Microservice}  & \textbf{application model}   & \textbf{composition to}   &  \textbf{Fog} & \textbf{Microservice} 
            \\
            & & \textbf{Management} & \textbf{Placement} & \textbf{Task} & \textbf{Computing} & & \textbf{Taxonomy} & \textbf{Specific} &  \textbf{Application} & \textbf{Composition}  & \textbf{to placement problem} &  \textbf{placement problem }   & \textbf{Platforms}   & \textbf{Workloads} \\
            
            &  & & & \textbf{Scheduling}
            & & & & \textbf{Taxonomy} & & & \textbf{formulation} & \textbf{formulation}   & &  \\
           
		   \hline
		   \cite{hong2019resource} & o & \checkmark & $\Delta$ & o & o & o & o & o & o & o & o & o & o & o    \\

		   \hline
		   \cite{jamil2022resource} & o & \checkmark & o & \checkmark & o & $\Delta$ & $\Delta$ & o & o & o & o & o & o & o    \\
              \hline
		   \cite{islam2022optimal} & $\Delta$ & o & \checkmark & o & o & \checkmark & \checkmark & o & o & o & o & o & o & o  \\
		   
		   \hline
		   \cite{mahmud2020application} & o & $\Delta$ & \checkmark & o & o & \checkmark & \checkmark & o & o & o & o & o & o & o  \\
		   
		   \hline
		   \cite{goudarzi2022scheduling} & o & o & $\Delta$ & \checkmark & o & \checkmark & \checkmark & o & o & o & o & o & $\Delta$ & o \\
		  
		   \hline
		   \cite{brogi2020place} & o & o & \checkmark & o & o &$\Delta$ & o & o & $\Delta$ & 0 & $\Delta$ & o & o & o   \\
		   \hline
		   \cite{salaht2020overview} & o & o & \checkmark & o & o & \checkmark & \checkmark & o & o & o & o & o & $\Delta$ & o \\
		   \hline
            \cite{varshney2020characterizing} & o & $\Delta$ & $\Delta$ & $\Delta$ &o &  \checkmark & \checkmark & o & o & o & o & o & o & o \\
              \hline
              \cite{joseph2019straddling} & \checkmark & o & $\Delta$ & o & $\Delta$ & 
              \checkmark & o & $\Delta$ & $\Delta$ & $\Delta$ &  o & o & o & o \\
              \hline
              \cite{razzaq2020systematic} & \checkmark & o & o & o & o & \checkmark & o & o & $\Delta$ & $\Delta$ & o & o & o& o \\ 
              \hline
                \cite{garriga2017towards} & \checkmark & o & o & o & o & \checkmark & o & $\Delta$ & $\Delta$ & $\Delta$ & o & o & o & o \\
              \hline
              \cite{neha2022systematic} & \checkmark & o & o & o & \checkmark & $\Delta$ & o & o & o & $\Delta$ & o & o & o & o \\
              \hline
              \cite{androvcec2019systematic} & \checkmark & o & o& o & \checkmark & o & o & o & o & o  & o & o & o & o  \\
              \hline
		   \textbf{Our} & \checkmark & o & \checkmark & o & $\Delta$  & \checkmark & o & \checkmark & \checkmark & \checkmark & \checkmark & \checkmark & \checkmark & \checkmark  \\
		   \hline	
	\end{tabular}
        \renewcommand{\arraystretch}{1.2}}
        \raggedright\footnotesize{\checkmark: Discussed, $\Delta$: Partially Discussed, o: Not Discussed}
\end{table*}



\section{Microservice Architecture}\label{appModelling}

\begin{figure}[h]
    \includegraphics[width=\linewidth]{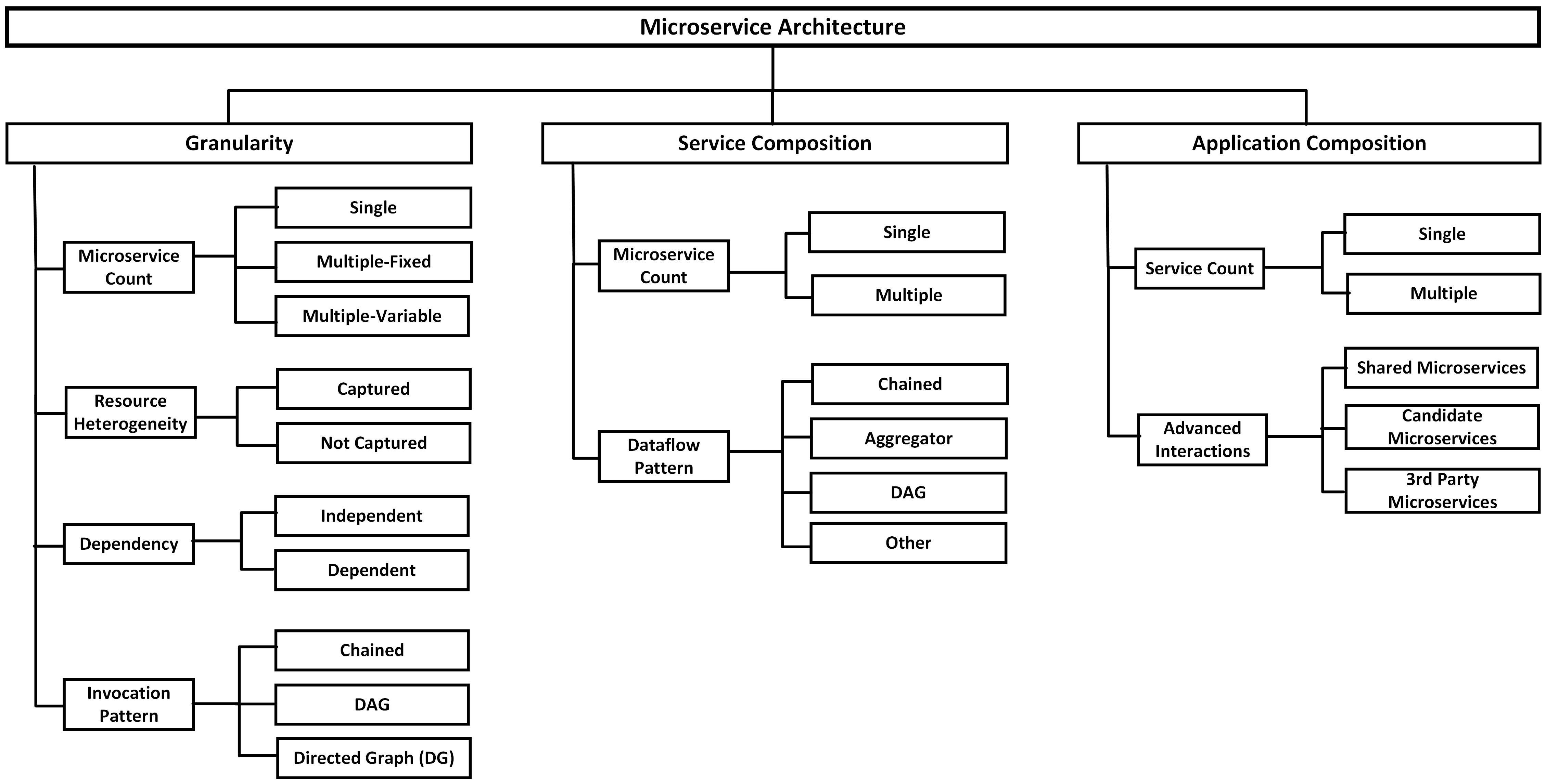}
    \caption{Taxonomy for modelling of Microservice Architecture for placement problem formulation}
    \label{fig:architectureTaxonomy}
    \vspace*{-5.5mm}
\end{figure}

To utilise the capabilities of MSA and overcome the challenges of the architecture within Fog environments, proper modelling of the applications is of vital importance so that the placement algorithms can capture all aspects of the placement problem. To this end, Fig. \ref{fig:architectureTaxonomy} presents the taxonomy for modelling applications based on MSA where we analyse microservices-based application modelling at multiple levels; granularity at microservice level (\textbf{\textit{Granularity}}), the composition of microservices at the service level (\textbf{\textit{Service Composition}}) and composition of services at the application level (\textbf{\textit{Application Composition}}). Table \ref{table:tax1Works} maps the existing works to the proposed taxonomy based on how each research work models the microservices-based applications.

\begin{table*}[h]
	\caption{Analysis of existing literature based on the taxonomy for modelling of Microservice Architecture}
	\label{table:tax1Works}
	\resizebox{\linewidth}{!}
	{\renewcommand{\arraystretch}{1.1}
	\begin{tabular}[htbp]{ |c|c|c|c|c|c|c|c|c|c|c|
	}
			\hline
			\multicolumn{1}{|c}{} & \multicolumn{4}{|c}{\textbf{Granularity}} &\multicolumn{2}{|c}{\textbf{Service Composition}} &\multicolumn{4}{|c|}{\textbf{Application Composition}}
			\\
			\cline{2-11}
			\textbf{Work} & \textbf{Microservice}  & \textbf{Resource} 
			& \textbf{Dependency} & \textbf{Invocation} & \textbf{Microservice}  & \textbf{Dataflow}
			& \textbf{Service}  & \multicolumn{3}{c|}{\textbf{Advanced Interactions}} 
			\\
		    \cline{9-11}
			& \textbf{count} & \textbf{Heterogeneity} & & \textbf{pattern} & \textbf{count} & \textbf{pattern} & \textbf{count} & \textbf{Shared} & \textbf{Candidate} & \textbf{3rd Party} 
			\\ 
			\hline
			\cite{filip2018microservices} & Multi-V & Captured ($P_C$) & Dependent & Ch & Multiple & Ch & NA & - & -&-\\
			\hline
			\cite{lera2018availability} & Multi-V & Captured (RU) & Dependent & DG & Multiple & ND & Single & - & - & - \\
			\hline
			\cite{guerrero2019lightweight} & Multi-V & Captured ($P_C$) & Dependent & DG & Multiple & NA & Multiple & - & - & -\\
			\hline
			\cite{faticanti2019cutting} & Multi-F & Captured (R,$P_C$,S) & Dependent & Ch & Multiple & Ch & Single & - & - & -  \\ 
			\hline
			\cite{pallewatta2019microservices} & Multi-V & Captured (R,$P_C$,S,B) & Dependent & DAG & Multiple & NA & Multiple & \checkmark & - & -  \\
           \hline 
           \cite{guerrero2019evaluation} & Multi-V & Captured (RU) & Dependent & DAG & NA & NA & NA & - & -& - \\
          \hline 
          \cite{faticanti2020throughput} & Multi-V & Captured (R,$P_C$,S,B) & Dependent & DAG & Multiple & DAG & Single & -  &  -  & - \\
          \hline
          \cite{wang2019delay} & Single & NC & Independent & NA & Single & NA & Single &  - & - & - \\
          \hline
          \cite{paul2020crew} & Mult-V & Captured ($P_C$, R) & Dependent & ND & ND & ND & ND & - & - & - \\
          \hline
          \cite{samanta2020dyme} & ND & Captured (RU) & Independent & NA & ND & NA & ND & - & - & - \\
          \hline
          \cite{fang2020iot} & Multi-V & Captured ($P_C$) & Dependent & ND & Multiple & Ch & Multiple & \checkmark & - & - \\
          \hline
          \cite{abdullah2020predictive} & Single & Captured ($P_C$) & Independent & NA & Single & NA & Single & - & - & - \\
          \hline
          \cite{faticanti2020deployment} & Multi-V & Captured(R,$P_C$,$P_G$,S) & Dependent & DAG & Multiple & DAG & ND & - & - &-\\
          \hline
          \cite{zhao2019distributed} & Multi-V & Captured ($P_C$) & Dependent & Ch & Multiple & Ch & Single & - & \checkmark & - \\
          \hline
          \cite{deng2020optimal} & Multi-V & Captured (R,$P_C$,S) & Dependent & Ch & Multiple & Ch & Single & - & -& - \\
          \hline
          \cite{xu2020service} & Multi-V & NC & Dependent & Ch & Multiple & Ch & ND & - & -& - \\
          \hline
          \cite{huang2020ant} & Multi-V & Captured ($P_C$) & Dependent & Ch & Multiple & Ch & ND & - & - & - \\
          \hline
          \cite{lei2020heuristic} & Multi-V & Captured ($P_C$) & Dependent & DAG & Multiple & DAG & Multiple & - & - & \\
          \hline
          \cite{armani2021cost} & Multi-V & Captured ($P_C$,R,S) & Dependent & DG & Multiple & DG & ND & - & -& -   \\
         \hline
         \cite{xu2021adaptive} & Single & NC & Independent & NA &  Single & NA & Single & - & -& - \\
         \hline
         \cite{baburao2021load} & Single & NC & Independent & NA & Single & NA & ND & - & -& -\\
         \hline
         \cite{herrera2021optimal} & Multi-V & Captured (R) & Dependent & Ch & Multiple & Ch & Multiple & \checkmark & - & - \\
         \hline
         \cite{he2021online} & Multi-V & Captured (R,$P_C$,S) & Dependent & DAG & Multiple & Ch, Ag, H & Multiple & \checkmark & - & - \\
         \hline
         \cite{watanabe2021afc} & Multi-V & Captured ($P_C$, R) & Dependent & DG & Multiple & DG & Single & - & \checkmark & \checkmark \\
         \hline
         \cite{fu2021qos} & Multi-V & Captured ($P_C$, R, B) & Dependent & DAG & Multiple & DAG & ND & - & - & - \\
         \hline
         \cite{alencar2021dynamic} & Single & Captured (R,B) & Independent & NA & Single & NA & ND & - & -& - \\
         \hline
         \cite{fu2021adaptive} & Multi-V & Captured ($P_C$, R, B) & Dependent & DAG & Multiple & DAG & ND & - &- & - \\
          \hline
         \cite{pallewatta2022qos} & Multi-V & Captured (R,$P_C$,S) & Dependent & DAG & Multiple & Ch, Ag, H & Multiple & \checkmark & - & -\\
          \hline
          \cite{guo2022joint} & Multi-V & NC & Dependent & DAG & Multiple & DAG & ND & - & -& - \\
          \hline
          \cite{lv2022microservice} & Multi-V & Captured ($P_C$,R) & Dependent & NA & Multiple & UWG & NA & - & - & - \\
          \hline
          \cite{zhang2022astraea} & Multi-V & Captured ($P_G$) & Dependent & DAG & Multiple & DAG & Single & - & -& -\\
          \hline
          \cite{kaur2022latency} & Multi-V & Captured (RU) & Dependent & NA & Multiple & DG & ND & - & - & - \\
          \hline
	\end{tabular}
	\renewcommand{\arraystretch}{1}}
	\raggedright\footnotesize{Mult-F:Multiple-Fixed, Mult-V:Multiple-Variable, Ch:Chain, Ag:Aggregator, H:Hybrid, DAG:Directed Acylclic Graph, DG:Directed Graph, UWG: Undirected Weighted Graph, R:RAM, $P_C$:Processing power (CPU), $P_G$:Processing power (GPU), S:Storage, B:Bandwidth, RU:Resource Units, NC:Not Captured, ND: Not Defined, NA:Not Applicable}
	\vspace*{-4.5mm}
\end{table*}

\subsection{Granularity}

Granularity is one of the most important and challenging aspects of MSA. The fine-grained nature of the microservices allows application services to be depicted as a collection of small components communicating together to perform a certain end-user-requested service. While this allows QoS-improved placement within resource-constrained Fog devices and dynamic movement between federated Fog-Cloud environments following the Osmotic computing paradigm, it introduces complexities in interaction patterns among the microservices. Thus, when modelling the application placement problem, the level of granularity should be captured accurately to overcome the challenges while utilising the advantages introduced by the granular design.

Hence, we analyse the microservice granularity within IoT applications as follows:
\begin{enumerate}
    \item \textit{Microservice count}: MSA decomposes the application into a set of microservices, increasing the complexity of the placement problem as the number of microservices increases. Thus, existing works capture different levels of granularity based on the number of microservices in the modelled IoT applications. Each application modelled in \cite{wang2019delay, alencar2021dynamic, xu2021adaptive, baburao2021load} consists of a single microservice such that the microservice is designed to perform a specific task requested by the end-user. As an example scenario, \cite{wang2019delay} introduces an object detection application used by autonomous cars for the detection of other vehicles, pedestrians, road signs etc., that consists of a single microservice for object detection service. To avoid the complexities introduced by having a large number of interconnected microservices, \cite{faticanti2019cutting} simplifies the placement problem by designing the placement algorithm to handle applications with a fixed number of microservices. \cite{faticanti2019cutting} proposes the placement algorithm for applications consisting of two microservices: a high throughput microservice that receives data and pe-process it to reduce the throughput, and a low throughput microservice which process the data sent from the first microservice. Works such as \cite{faticanti2020throughput, zhao2019distributed, pallewatta2022qos, lera2018availability, guerrero2019evaluation} remove this constraint and model the application as a collection of any number of microservices, thus providing robust placement algorithms that capture the problem-domain dependent granularity levels of MSA more accurately.
    \item \textit{Resource Heterogeneity}: One of the advantages of decomposing applications into microservices is to achieve functional separation where the application is divided into separate modules following the "separation of concern" design pattern. This results in the separation of microservices based on their resource requirements as well (i.e., CPU, GPU, RAM, storage etc.). It's especially advantageous in Edge/Fog environments where resources are heterogeneous (i.e., Raspberry Pi \footnote{https://www.raspberrypi.com/products/raspberry-pi-4-model-b/specifications/}, Jetson Nano \footnote{https://developer.nvidia.com/embedded/jetson-nano}, Dell PowerEdge XR12 \footnote{https://www.dell.com/en-au/work/shop/cty/pdp/spd/poweredge-xr12/aspexr12\_vi\_vp}, Lenovo ThinkEdge SE50 \footnote{https://psref.lenovo.com/syspool/Sys/PDF/ThinkEdge/ThinkEdge\_SE50/ThinkEdge\_SE50\_Spec.pdf}, etc.) and resource-constrained unlike in Cloud environments. \cite{faticanti2019cutting, pallewatta2022qos, he2021online} demonstrate this by modelling microservices within the same application to have heterogeneous resource requirements in terms of multiple resource parameters such as RAM, CPU, storage and bandwidth. \cite{faticanti2020deployment, zhang2022astraea} extend this to include GPU as well, where microservices with GPU requirements may have to be moved to different Fog locations or Fog service providers based on the GPU availability.
    While the above works represent resource requirements as a vector, some works like \cite{guerrero2019evaluation, lera2018availability} simplify the representation by introducing the scalar parameter "Resource Units", with the possibility of extending it to include multiple resource types.
    \item \textit{Dependency among microservices}: Microservices are developed as independently deployable units with well-defined business boundaries such that their functionality is exposed to the outside through open interfaces. This allows microservices within applications to communicate easily with each other to create composite services. \cite{wang2019delay, samanta2020dyme, abdullah2020predictive} represent microservices as independent entities without any interconnections among them. In contrast, works such as \cite{filip2018microservices, lera2018availability, pallewatta2019microservices} model them to have dependencies, where microservices communicate with each other through lightweight communication protocols such as REST APIs and message brokers, creating a plethora of IoT services. 
    \item \textit{Invocation pattern}: The higher level of openness supported by microservices results in different invocation patterns among them, where client microservices invoke other microservices to perform functions required to complete the composite services. \cite{deng2020optimal, filip2018microservices, zhao2019distributed} models the invocation relationship as a chained pattern. \cite{he2021online, pallewatta2022qos, guo2022joint} use a more general representation of microservice invocation by modelling it using DAG representation. \cite{guerrero2019lightweight} and \cite{ lera2018availability} model the invocation as directed graphs where the interactions between microservice are depicted using many-to-many consumption relationships.
\end{enumerate}

\subsection{Service Composition}

With granularity comes the concept of composite services, where microservices interact to create services that perform a certain task and provide an output to the user. In this section, we analyse and categorise aspects related to service composition.

\begin{enumerate}
    \item \textit{Microservice count}: The granularity of the microservices creates end-user services with varying numbers of microservices: "atomic services" consisting of a single microservice and "composite services" that consist of multiple interconnected microservices. 
    \cite{wang2019delay, abdullah2020predictive} represent each service by a single microservice that receives requests from the end-user front-end, completes a task and provides the result back to the user without interacting with any other microservices. Other works like \cite{deng2020optimal,guo2022joint, pallewatta2022qos} model services with multiple microservices that interact together to perform tasks. \cite{deng2020optimal} describes a smart city application with a smart policing service used by the police to identify suspects where the service is a composition of three microservices, and \cite{pallewatta2019microservices, pallewatta2022qos} model a smart healthcare application with an emergency notification service where multiple microservices interact to detect abnormalities in ECG data streams and raise emergency alarms in real-time.
    
    \item \textit{Dataflow Pattern}:
 Due to different interaction patterns among microservices, the data flow within composite services can take many forms. \cite{filip2018microservices, faticanti2019cutting, fang2020iot} consider the chained composition of microservices, whereas \cite{he2021online, pallewatta2022qos} models the services considering chained, aggregator and hybrid dataflow patterns to create composite services. Works such as \cite{faticanti2020deployment, lei2020heuristic} model dataflow among microservices as DAGs, assuming the absence of cyclic data flows while \cite{armani2021cost, watanabe2021afc} include cyclic dataflows and represent the dataflows using DGs.  \cite{lv2022microservice} models the interaction pattern using an Undirected Weighted Graph (UWG) where edges are represented using interaction weights irrespective of the direction of communication.\end{enumerate}

\subsection{Application Composition}
As the capabilities of IoT applications improve rapidly, they quickly evolve into complex applications covering large business domains due to the flexibility of design and development using MSA. We analyse the aspects related to this as follows: 
\begin{enumerate}
    \item \textit{Service count}: With the increase in connected devices and the generation of diverse data, IoT applications have evolved to provide many services to users. Thus, microservices-based IoT applications are modelled as a composition of one or more services. \cite{faticanti2019cutting, wang2019delay} define applications with single services, whereas \cite{guerrero2019lightweight, fang2020iot, lei2020heuristic} model applications with multiple end-user services with heterogeneous QoS requirements. \cite{pallewatta2019microservices, pallewatta2022qos} model a smart health care application with two primary services; emergency alarm generation service with stringent latency requirements and a latency tolerant long-term analysis service. Modelling the applications with multiple services allows the placement problem to capture competing QoS requirements within applications and paves the way to propose placement policies that can utilise Fog-Cloud resources more efficiently.
    
    \item \textit{Advanced Interactions}: Due to the fine-grained nature of the microservices, along with well-defined functional boundaries and lightweight communication methods, advanced interactions among microservices are possible. \cite{pallewatta2022qos} models applications where some microservices (i.e., feature extraction, data cleaning etc.) are part of multiple composite services that have heterogeneous latency requirements. \cite{zhao2019distributed} models the services to have candidate microservices (i.e., online payment gateways) depending on the payment method used by the user. Thus, the application is modelled to have alternative data paths. \cite{alencar2021dynamic} handles the concept of having alternative paths by defining control structures for each composite service through conditional branching within some of the microservices. Another advanced capability of microservices is using third-party microservices through the use of APIs, which is modelled in the \cite{alencar2021dynamic} where composite services are created by combining microservices developed by multiple developers, and provided as services by multiple computing service providers.
\end{enumerate}

\subsection{Research Gaps}
Based on the analysis of existing works presented in Table \ref{table:tax1Works}, we have identified the following gaps related to microservices-based application modelling.
\begin{enumerate}
    \item Existing works show shortcomings in several aspects in capturing the granularity of the MSA, such as lack of use in generalised invocation patterns (i.e., directed graphs) and capturing resource heterogeneity in different microservices within the same application (i.e., use of GPU, databases etc.). Many works use a chained invocation of microservices without considering complex dependency patterns that can occur due to the openness of microservice design. For capturing resource heterogeneity,  most works consider one or more parameters such as CPU, RAM and storage. However, GPU or TPU requirements are scarcely considered. With the rise of EdgeAI workloads in IoT, such parameters and the constraints imposed by them play an important role in the placement logic.
    \item In service composition, the data flow pattern is not adequately defined and utilised in the placement process of the majority of the works. Many works define chained data flow or acyclic data flows, disregarding cyclic data flows, which affect the end-to-end latency calculations of the services.
    \item Under application composition, most works model each application to have a single composite service. Thus heterogeneity in QoS requirements between different composite services within the same IoT application is not appropriately captured. This also hinders the application from having complex interaction patterns and data flow patterns within the applications.
    \item In modelling the applications, most works do not consider complex interaction patterns of microservices, including shared microservices among different composite services, candidate microservices and third-party microservices. The placement of shared microservices has to consider multiple, heterogeneous composite services that they are part of and place them so that all non-functional requirements of the services are satisfied under resource contention. For efficient placement of the candidate services, knowledge of their demand and usage is vital. The use of third-party microservices in applications results in security, reliability and performance challenges  (i.e., latency, availability, transaction cost etc.)  that are out of the control of the application developers, which need to be accounted for during the placement phase to improve performance. Most of the works that consider these only analyse the effect of a single pattern, thus failing to capture the impact of multiple ones.
\end{enumerate}

\section{Application Placement Policy}\label{tax2}

\begin{figure*}[!ht]
    \includegraphics[width=0.6\linewidth]{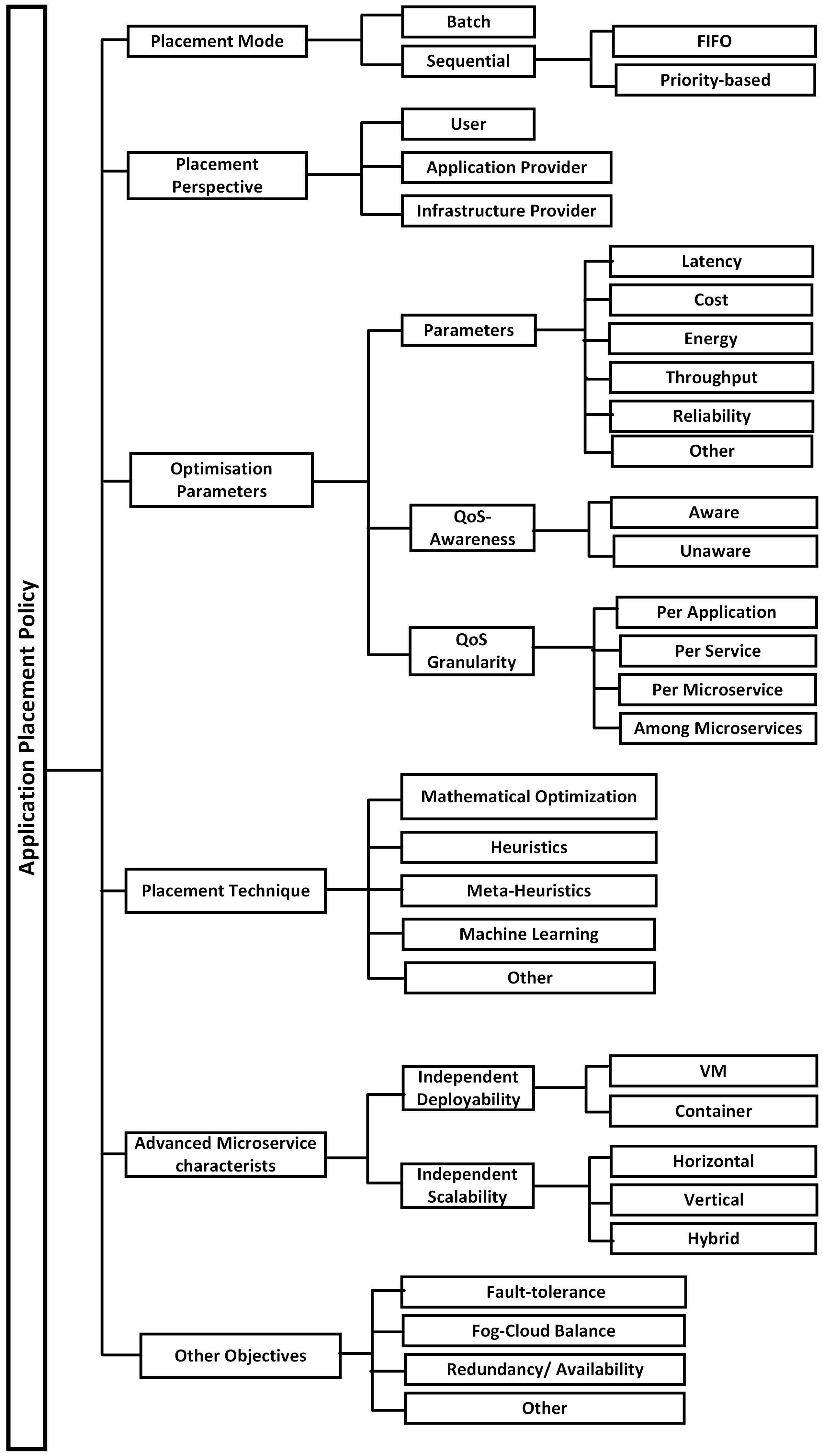}
    \caption{Taxonomy for placement policies designed for microservices-based applications}
    \label{fig:placementTaxonomy}
    \vspace*{-4.5mm}
\end{figure*}

MSA introduces novel aspects (i.e., QoS granularity, scalability, lightweight/ independent deployment,  etc.) that can be utilised for better placement of the IoT applications while also giving rise to novel challenges (i.e., microservice dependencies, interactions, cascading failures etc.). We consider these MSA-specific effects in addressing the application placement problem and create a novel taxonomy as shown in Fig. \ref{fig:placementTaxonomy}. Current works are mapped to the taxonomy to identify gaps and possible improvements (see Table \ref{table:tax2Works}).

\subsection{Placement Mode}
Placement mode represents the number of placement requests processed by the placement engine during each execution of the placement algorithm. Placement modes are categorised into two groups: sequential mode, where the placement algorithm queues the placement requests to process one after the other, and batch mode, where a set of applications are considered for placement simultaneously. In the works such as \cite{pallewatta2019microservices, deng2020optimal, wang2019delay}, the placement engine uses a First In First Out (FIFO) queue to store and process application placement requests sequentially. \cite{faticanti2019cutting, faticanti2020throughput} prioritise the applications in the queue based on their resource requirements, whereas \cite{lera2018availability} and \cite{xu2021adaptive} prioritise them based on the deadlines of the applications and order them for sequential placement. Placement engines in \cite{guerrero2019evaluation, pallewatta2022qos, lei2020heuristic} are designed using meta-heuristic algorithms (i.e., genetic algorithm, particle swarm optimisation) to handle the complexities of the batch placement and navigate the larger solution space successfully.  \cite{faticanti2020deployment} employs a batch placement policy based on a heuristic algorithm to maximise the placement of IoT applications within a multi-domain federated Fog ecosystem under locality constraints imposed on microservices. Some works, such as \cite{huang2020ant, abdullah2020predictive, xu2020service} do not properly define placement mode and carry out the evaluations using the placement of a single application.

\subsection{Placement Perspective}
The placement perspective is identified based on the optimization parameters/objectives considered during the placement and from whose viewpoint they are addressed from. \cite{faticanti2019cutting, faticanti2020throughput, armani2021cost} define the placement problem from the perspective of the Fog infrastructure provider, thus aiming to maximize the IaaS provider's revenue. Other works such as \cite{deng2020optimal, zhao2020distributed, fang2020iot} define the placement problem from the application provider perspective, where application providers expect the satisfaction of non-functional requirements (i.e., latency, throughput, reliability etc.) of the services provided by their application while ensuring budget constraints related to Fog-Cloud deployment. \cite{guerrero2019lightweight, filip2018microservices} address this from a user perspective where the user sends a request for an application or service along with performance requirements, and the placement algorithm ensures the availability of the requested service under requested constraints (i.e., latency, throughput etc.) to satisfy the user expectations. 

\afterpage{
\begin{landscape}
\begin{table}[!htbp]
	\caption{Analysis of existing literature based on the taxonomy for application placement policy}
	\label{table:tax2Works}
	\resizebox{\linewidth}{!}
	{\renewcommand{\arraystretch}{1}
	\begin{tabular}[htbp]{ |c|c|c|c|c|c|c|c|c|c|
	}
			\hline
			\multicolumn{1}{|c}{\textbf{Work}} &
			\multicolumn{1}{|c}{\textbf{Placement}} &
			\multicolumn{1}{|c}{\textbf{Placement}} &
			\multicolumn{3}{|c}{\textbf{Placement Parameters}} &
			\multicolumn{1}{|c}{\textbf{Placement}} &\multicolumn{2}{|c}{\textbf{Adv $\mu$service characteristics}}&
			\multicolumn{1}{|c|}{\textbf{Other Objectives}}\\
			\cline{4-6} \cline{8-9} 
			& \textbf{Mode}  & \textbf{Perspective} & \textbf{Parameters} & \textbf{QoS-awareness}  & \textbf{QoS Granularity} &
			 \textbf{Techniques} & \textbf{Ind. Deployability} & \textbf{Ind. Scalability} & 
			\\
			\hline
			\cite{filip2018microservices} & Seq-FIFO & UP  & Latency & No & per microservice/s (Latency) & Heuristic - Greedy & VMs & NC & - \\
			\hline
			\cite{lera2018availability} & Seq-P & AP & Latency, Availability & Yes (Latency) & per application (Latency) &  Heuristic & Containers & H & - \\
			\hline
			\cite{guerrero2019lightweight} & Seq-ND & UP  & Latency & No & per service (Latency) &  Heuristic & Containers & H & F-C Balance (D) \\
			\hline
			\cite{faticanti2019cutting} & Seq-P & IP  & Revenue, Latency  & Yes  & among microservices (Throughput) &  Heuristic-Greedy & Containers & NC & F-C Balance (S) \\ 
			& & & Throughput & (Latency, Throughput) &  per application (Latency) & & & &\\
			\hline
			\cite{guerrero2019evaluation} & Batch & AP & Latency, $RU_C$ & No & ND &  Meta-Heuristic & Containers & H & Service Spread \\
			\hline 
			\cite{pallewatta2019microservices} & Seq-FIFO & UP  & Latency & Yes & per service (Latency) &  Heuristic & Containers & H+V & F-C Balance (S) \\
			\hline
			\cite{huang2020ant} & ND & AP & Latency, Cost & No & per service &  Meta-Heuristic & Containers & H+V & Redundancy \\
			\hline
			\cite{paul2020crew} & Seq-ND & AP & Cost, Reliability & Yes (All) & per application (All) &  Meta-Heuristic & ND & H & - \\
			& & & Latency & & & & &  & \\
			\hline
			\cite{xu2020service} & ND & AP & Latency, Energy & No & NC &  Meta-Heuristic & Containers & H & Redundancy \\
		    \hline	
		    \cite{abdullah2020predictive} & ND & UP & Latency & Yes (Latency) & per microservice (Latency) &  Machine Learning & Containers & H & Proactive Scaling \\
			\hline 
			\cite{samanta2020dyme} & Batch & UP & Energy, Latency & Yes (Latency) & per microservice (Latency) &  Mathematical Programming & Containers & H & Fault-tolerance \\
			& & & Cost &  & & (Lagrangian Multiplier) & && \\
			\hline
			\cite{lei2020heuristic} & Batch & AP & Latency, Cost & Yes (Latency) & per service (Latency) &  Meta-Heuristic & ND & NC & Fault-tolerance \\
			\hline
			\cite{fang2020iot} & Seq-ND & AP & Energy, Latency & No & NC &  Meta-Heuristic & VMs & NC & - \\
			\hline
			\cite{faticanti2020deployment} & Batch & IP &   Revenue, Latency, & Yes  & among microservices &  Heuristic & Containers & NC & Locality-awareness \\
			& &  &Throughput & (Latency, Throughput) & (Latency, Throughput) &  & & &\\
			\hline
			\cite{faticanti2020throughput} & Seq-P & IP  & Revenue, Latency & Yes  & among microservice (Throughput) &  Heuristic-Greedy & Containers & NC & F-C Balance (D) \\
			& & & Throughput & (Throughput, Latency) & per application (Latency) &  &&& \\
			\hline
			\cite{wang2019delay} & Seq-FIFO & AP/UP & Latency, Cost & Yes (Latency) & per application (Latency) &  Reinforcement Learning & Containers & NC & Mobility-awareness \\
			\hline
			\cite{armani2021cost} & Seq-ND & IP & Latency, Throughput & Yes (All) & among microservices (All) &  Heuristic & Containers & H & F-C Balance (D) \\
			\hline
			\cite{xu2021adaptive} & Seq-P & UP+AP & Latency, Cost & Yes (Latency) & per microservice (Latency) &  Heuristic - Greedy & Containers & NC & F-C Balance (D) \\
			\hline
			\cite{he2021online} & Seq: ND & AP & Latency, Cost & Yes (Cost) & among microservices (Latency) &  Heuristic - Greedy & Containers & H & -\\
			& & & & & per application (Cost) & &  & &\\
			\hline
			\cite{alencar2021dynamic} & Seq-ND & AP & Latency, $RU_C$ & Yes (Throughput) & per microservice &  Analytical Hierarchy Process  & Containers & NC & - \\
			& & & Throughput & & (Throughput, Latency) & (AHP) & & &\\
			\hline
			
			\cite{fu2021qos} & ND & IP+UP & Latency, Throughput & Yes & per application &  Approximate Algorithm & Containers & H & Load-awareness \\
			& & & Cost, $RU_C$, $RU_N$ & (Latency, Throughput) & (Latency, Throughput) &  + Deep Reinforcement Learning & & & Resource contention \\
			\hline
			 
			\cite{pallewatta2022qos} & Batch & AP & Latency, Cost & Yes (All) & per service (All) &  Meta-Heuristic & Containers & H+V & F-C Balance (D) \\
			& & & Throughput, $RU_C$, $RU_N$ &  & & & & & \\
			\hline
			
			\cite{guo2022joint} & Batch & UP & Latency, $RU_N$ & Yes (Latency) & per service (Latency) &  Meta-Heuristic & Containers & H & Location-awareness \\ \hline
			
			\cite{lv2022microservice} & ND & AP & Latency, $RU_C$ & No & NC &  Deep Reinforcement Learning & Containers & H & Load-awareness \\
			& & & & &  & + Heuristic & & & \\
			\hline

			\cite{deng2020optimal} & Seq-FIFO & AP & Cost, Latency & Yes (Latency) & per service (All) &  Mathematical Programming & Containers & H+V & - \\
			& & & & &  & (Branch and Bound) & & &\\
			\hline
			\cite{zhao2020distributed} & Seq-FIFO & AP & Latency & No & per service/application &   Monte carlo +  & Containers & H & Availability \\
			& & & & & (Latency)  & Meta-heuristic & & & \\
		\hline	
		\cite{herrera2021optimal} & Batch & IP & Latency & No & per service (Latency) &  Mathematical Programming & ND & H+V & Optimal routing \\
			& & & & &  & (Gurobi MILP solver) & & & \\
			\hline
			\cite{watanabe2021afc} & Seq-ND & UP & Latency, Throughput & Yes & among microservices &  Mathematical Programming & VMs & H & Redundancy \\
			& & & & (Latency, Throughput) & (Latency, Throughput)  & & & & \\
			\hline
			\cite{fu2021adaptive} & ND & IP+UP & Latency, Throughput & Yes & per application &  Approximate Algorithm & Containers & H & Load-awareness \\
			& & & Cost, $RU_C$, $RU_N$ & (Latency, Throughput) & (Latency, Throughput) &  + Deep Reinforcement Learning & & & Resource contention \\
            \hline
			\cite{zhang2022astraea} & ND & UP & Latency, Throughput & Yes (Latency) & per service (Latency) &  Heuristic-Greedy & ND & H & Resource contention \\
                \hline
                \cite{mortazavi2022discrete} & Batch & IP + UP & Latency, Energy & Yes & among microservices & Meta-heuristic & Containers & H & Fault-tolerance \\
                & & & Throughput & (Latency, Throughput) & (Latency, Throughput) & & & & \\
			\hline
             \cite{kaur2022latency} & Batch & UP & Latency & No & among microservices & Meta-Heuristic & Containers & NC & - \\
             \hline
	\end{tabular}
	\renewcommand{\arraystretch}{1}}
	\raggedright\footnotesize{FIFO:First-In-First-Out, P:Prioritised, UP:User Perspective, AP:Application provider Perspective, IP:Infrastructure provider Perspective, $RU_C$:Computation Resource Utilisation, $RU_N$:Network Resource Utilisation, ND:Not Defined, NC:Not Considered, H:Horizontal, V:Vertical, F-C Balance(D):Dynamic Fog-Cloud Balance, F-C Balance(S):Static Fog-Cloud Balance}
\end{table}
\end{landscape}
}

\subsection{Placement Parameters}
In this section, we analyse the characteristics of the parameters considered by the placement policy.
\begin{enumerate}
    \item \textit{Parameters: } IoT application services have multiple heterogeneous QoS parameters (i.e., latency \cite{watanabe2021afc, herrera2021optimal, zhao2020distributed, deng2020optimal, lv2022microservice}, cost \cite{deng2020optimal, pallewatta2022qos, xu2021adaptive}, throughput \cite{watanabe2021afc, alencar2021dynamic, armani2021cost}, energy \cite{xu2020service}, reliability \cite{paul2020crew} etc.) negotiated with the Edge/Fog infrastructure provider in the form of Service Level Agreements (SLA). Moreover, placement decisions made from the infrastructure provider perspective consider the maximisation of the revenue of the IaaS provider. \cite{faticanti2019cutting}, and \cite{faticanti2020throughput} formulate the placement problem to maximise the revenue of the Fog provider, whereas \cite{faticanti2020deployment, armani2021cost} consider a federate Fog environment where each provider focuses on minimising the total deployment cost while minimising the number of resources rented from external Fog infrastructure providers and the Cloud. Resource utilisation is another parameter considered in current research where \cite{guerrero2019evaluation, lv2022microservice, alencar2021dynamic} consider optimisation of computation resource usage, \cite{guo2022joint} optimises network resource usage, while \cite{pallewatta2022qos, fu2021adaptive} consider both. Placement policies focus on optimising one or more of the above-mentioned QoS parameters. \cite{filip2018microservices, pallewatta2019microservices, guerrero2019lightweight} formulate the placement problem considering the satisfaction of a single parameter and focus on minimising the latency of the services deployed within Fog environments. Other works like \cite{lei2020heuristic, paul2020crew, samanta2020dyme} focus on multiple parameters to reach a trade-off between conflicting parameters, such as latency, cost, energy etc., by formulating the placement problem as a multi-objective optimisation problem. 
    
    \item \textit{QoS-awareness: } Reduction of service latency and core network congestion are two of the main objective of Edge/Fog computing paradigms. As a result, \cite{guerrero2019lightweight, huang2020ant, zhao2020distributed} aim to place the microservices as close as possible to the user to reduce the overall latency of the application. While this is done in a QoS-unaware manner, assuming that all services require low latency, other research works such as \cite{paul2020crew, faticanti2020deployment, pallewatta2022qos, guo2022joint} consider the QoS requirements of the IoT services before placing them. Such approaches highlight one of the main limitations of Edge/Fog computing which is its resource-constrained nature compared to the Cloud data centres. Thus, they aim to capture the QoS heterogeneity of the IoT services and prioritise them based on their QoS requirements (i.e., stringent latency requirements over latency-tolerant services \cite{guo2022joint, faticanti2020deployment}, stringent budget constraints over higher budget availability \cite{paul2020crew, pallewatta2022qos}, throughput-aware placement/scaling of microservices \cite{faticanti2019cutting, pallewatta2022qos}, etc.) and achieve a proper balance between Fog and Cloud resource usage in a QoS-aware manner. MSA further enhances this behaviour due to the ease of moving independently deployable microservices across Fog and Cloud data centres dynamically. 
    
    \item \textit{QoS Granularity: } Because of the fine-grained nature of the microservices and their possible composition patterns, QoS parameters are defined at different levels of granularity. Among existing works, QoS parameters are defined at three primary levels: microservice level, composite-service level and application level. \cite{faticanti2019cutting, faticanti2020throughput} define the throughput requirements at the microservice level, where it is presented as the bandwidth requirement between interacting microservices, whereas \cite{filip2018microservices, abdullah2020predictive, samanta2020dyme} define latency requirements per each microservice. Several works such as \cite{guerrero2019lightweight, huang2020ant, pallewatta2022qos} define latency, throughput, cost etc., per each composite service where microservices-based applications can consist of one or more such services. \cite{faticanti2020throughput, wang2019delay} define latency requirement per each application, assuming that the application performs a single function/service requested by the end-user. Most of the works define all QoS parameters at a single level, whereas \cite{faticanti2020throughput, faticanti2019cutting} uses different granularity levels based on the parameter (i.e., throughput at the microservice level and latency at the application level). 
    
\end{enumerate}

\subsection{Placement Technique}

Different placement techniques are chosen to implement the placement policies based on the complexity of the modelled microservices-based applications, placement mode, optimisation parameters and their granularity, and the dynamism of the considered scenario (in terms of distributed resources, workload, failures etc.). Works such as \cite{samanta2020dyme, deng2020optimal, herrera2021optimal, watanabe2021afc} use mathematical programming to solve microservices-based application placement in Fog: \cite{herrera2021optimal} formulates the problem using Mixed Integer Linear Programming (MILP) and solve it using Gurobi MILP solver \footnote{https://www.gurobi.com/products/gurobi-optimizer/}, \cite{samanta2020dyme} uses the Lagrangian multipliers to solve the optimisation problem under multiple constraints including network QoS, price, resource usage etc. Several works, including \cite{filip2018microservices, guerrero2019lightweight, pallewatta2019microservices} propose heuristic placement algorithms to optimise a single placement parameter such as latency. Heuristic approaches are proposed by works such as \cite{lera2018availability, faticanti2019cutting, faticanti2020deployment} that consider multiple parameters as well. Here, \cite{faticanti2019cutting}, and \cite{faticanti2020deployment} formulate the placement problem as an Integer Linear Programming problem to solve it using greedy heuristic algorithms. \cite{guerrero2019evaluation, lei2020heuristic, guo2022joint} that try to handle batch placement scenarios and multiple placement parameters, propose placement policies based on evolutionary meta-heuristic algorithms such as Genetic Algorithm (GA), Particle Swarm Optimisation (PSO), Ant Colony Optimisation (ACO) to navigate large solution space efficiently. With increased computation power available for placement engines, algorithms are moving towards Machine Learning (ML): \cite{abdullah2020predictive} uses ML-based forecasting models to make predictive auto-scaling decisions, and \cite{wang2019delay, lv2022microservice, fu2021adaptive} use Reinforcement Learning based approaches to tackle the highly dynamic nature of Edge/Fog environments and the microservices. Moreover, existing works use other techniques such as Analytical Hierarchy Process (AHP) \cite{alencar2021dynamic}, which is a powerful decision analysis method used to prioritise and balance multiple criteria, and computation algorithms such as Monte Carlo method \cite{zhao2020distributed} used to capture uncertainties in the placement problem introduced by the MSA (i.e., candidate microservices, 3rd party microservices, etc.).

\subsection{Advanced Microservice Characteristics}

The popularity of the MSA for deployment within highly distributed Fog environments stems from two primary microservices-related characteristics, their independently deployable and scalable nature.

\begin{enumerate}
    \item \textit{Independent Deployability:} Decomposing monolithic applications into a collection of microservices can improve deployment-related aspects such as fault tolerance and reliability due to the independently deployable nature of the microservices. This is further supported by containerisation, which provides a lightweight method for deploying microservices compared to the earlier used Virtual Machines(VMs). \cite{filip2018microservices} proposes a deployment scenario where all microservices of a user-requested service are mapped onto a single VM. The vast majority of existing works \cite{lera2018availability, guerrero2019evaluation, xu2020service, faticanti2020deployment} improve the deployment through the use of containers (i.e., Docker), which allows rapid deployment by providing microservices with lightweight, isolated environments. The use of containers enables fast deployment of microservices while mitigating single point of failure using the distributed placement of microservices. Works such as \cite{samanta2020dyme,pallewatta2019microservices} highlight the importance of using container technology due to their faster spin-up times and analyse resultant performance improvements under dynamic conditions. \cite{deng2020optimal, pallewatta2022qos} consider the cost of the containerised microservices by adapting the pricing models used in container platforms provided by the commercial Cloud providers (i.e., Amazon Fargate, Azure Containers).
    
    \item \textit{Independent Scalability:} As microservices are independently deployable units with well-defined business boundaries, each microservice can be independently scaled to meet the throughput requirements. This is especially advantageous in Edge/Fog computing scenarios where devices are heterogeneous in their resource capacities, workloads also vary rapidly with the popularity of the services, mobility of the users etc. Scalability of the microservices is used in multiple ways where some works consider only horizontal scalability while others use a combination of horizontal and vertical scalability to satisfy throughput requirements. \cite{guerrero2019evaluation} uses horizontal scalability of the microservices to spread instances of each microservices uniformly across the Fog landscape to support distributed access by users.  \cite{pallewatta2019microservices, pallewatta2022qos} consider resource heterogeneity of the Fog devices and use a hybrid approach of vertical and horizontal scalability to meet the throughput requirement. Current works use the scalability of the microservices for multiple purposes. Above works focus on satisfying the overall throughput requirement while utilising resource-constrained Edge/Fog devices. Meanwhile, \cite{guo2022joint} uses horizontal scalability to incorporate location awareness, whereas \cite{zhao2020distributed, samanta2020dyme} use horizontal scalability to achieve redundancy which improves availability and fault-tolerance of the application. 
\end{enumerate}

\subsection{Other Placement Objectives}
One of the main advantages of using MSA for Fog application development is the ability to easily utilise both Fog and Cloud resources to meet application requirements. This concept is also put forth in Osmotic computing, which highlights the importance of dynamically moving microservices between Cloud and Fog and achieving an equilibrium such that the non-functional requirements of the services are met. In some works such as \cite{faticanti2019cutting, pallewatta2019microservices}, this is handled in a static manner where microservices to be placed on Cloud are predefined per each application based on the deadline requirement of the composite services each microservice belongs to. \cite{faticanti2020throughput, guerrero2019lightweight, armani2021cost, pallewatta2022qos} handle this in a dynamic way: \cite{faticanti2020throughput} partitions the application based on the throughput requirement to place microservices with higher throughput requirements in the Fog and rest in the Cloud, \cite{guerrero2019lightweight} analyses the popularity of the microservices based on the user requests to move less popular ones to the Cloud, \cite{armani2021cost} achieves balance by considering the cost of deployment in Fog and Cloud, \cite{pallewatta2022qos} formulates a multi-objective problem to achieve a trade-off between Fog device usage and network usage to achieve a balance between Fog and Cloud resource usage dynamically. 

Locality or location awareness is another aspect that comes with IoT applications due to data security and latency requirements. Sensitive IoT data (i.e., healthcare, security camera footage etc.) can have constraints on geographical locations for the processing, which requires certain microservices to be placed within certain regions or within certain Fog service providers in federated Fog computing environments. \cite{faticanti2020deployment} adds locality constraint to their problem formulation and handles it using a heuristic approach. With the distributed nature of the Edge/Fog resources and the users, location awareness can be used to minimise network resource usage and delay \cite{guo2022joint, guerrero2019evaluation}. \cite{guo2022joint} utilise this concept where the user's location is used to select the microservice instances such that the requests are routed to the closest instances. \cite{guerrero2019evaluation} introduces a metric called service spread which evenly distributes microservices across the Fog landscape to improve performance when users are uniformly distributed.

Modularity and scalability of microservices pave the way for efficient redundant placement to improve the availability and fault-tolerance of the services. \cite{zhao2020distributed} proposes a redundant placement policy for composite services consisting of chained microservices, considering the uncertainty of requests and heterogeneity of the resources. This work aims to reduce the outage time under failures through redundant placement of microservices. To address the challenge of cascading failures that occur in MSA, \cite{lei2020heuristic} proposes a fault tolerance method for applications deployed using the API gateway pattern where the API gateway acts as the access point for requests coming from the users. This work deploys API gateways within cache-enabled edge nodes so that data related to service requests can be cached proactively in case downstream microservices become unavailable due to failures until the microservice is redeployed using a reactive fault-tolerance approach.

Due to the horizontal scalability of microservices, proper load balancing and routing approaches are required to ensure performance requirements. To address this \cite{huang2020ant, xu2020service, herrera2021optimal} formulate the application placement problem as a combination of microservice replica placement and determining the optimal data flow path of composite services. \cite{huang2020ant} uses minimisation of service latency and cost (both computation and data transfer costs) to achieve the optimum level of service replication and identification of the request path that minimises latency of the composite services. \cite{xu2020service} considers latency and power consumption as optimisation parameters to achieve this. \cite{herrera2021optimal} extends this further by incorporating SDN controller placement as part of the problem, where SDN controllers are used for service discovery and determining data flow.

Deploying multiple lightweight containers onto the same Fog device can result in resource contention depending on the resource requirements of each container instance. To overcome this, \cite{fu2021qos, fu2021adaptive, zhang2022astraea} propose online algorithms to detect shared-resource contentions and afterwards dynamically adjust resource allocations or migrate microservice instances to other idle nodes to maintain the required level of performance. These works capture different types of resource contentions, including I/O \cite{fu2021adaptive}, GPU global memory bandwidth \cite{zhang2022astraea}, and computation capacity \cite{fu2021qos} contentions.

\subsection{Research Gaps}

Based on the analysis of existing works presented in Table \ref{table:tax2Works}, we identified the following gaps related to microservices-based application placement.
\begin{enumerate}
    \item QoS-granularity and QoS-awareness, together with proper placement mode selection, can improve performance, especially when microservices-based IoT applications are considered. As IoT applications grow in complexity to provide many services with heterogeneous QoS requirements, the granularity of the microservices paves the way for per-service QoS definitions. Together with batch placement or sequential placement with QoS-aware prioritisation, this approach is rarely explored in existing works.
    The following shortcomings are apparent in existing works; 
\begin{itemize}
    \item QoS-granularity: Most of the works define QoS requirements among interacting microservices (i.e., latency and throughput requirements between two microservices) which becomes less feasible due to composite services with complex interaction patterns and their agile evolution as applications evolve. As a result, existing works scarcely capture the QoS heterogeneity of the composite services within the same application. Thus, MSA-related scenarios like competing requirements that exist due to shared microservices are not handled.
    \item QoS-awareness: Many works do not define QoS requirements and capture the heterogeneity but try to minimise overall latency, cost etc. This approach hinders the proper utilisation of limited Edge/Fog resources and limits achieving equilibrium between Fog-Cloud resource usage.
\end{itemize}
    The above gaps are tightly coupled with how the microservices-based application is modelled, and overcoming them requires the proper capturing of MSA as described in Section \ref{appModelling}.
    \item In existing works, there's scope for incorporating advanced microservice characteristics such as the independently deployable and scalable nature of the microservices. Although many of the works use container technology for microservice deployment, they lack proper utilisation of the fast spin-up time, lightweight deployment capabilities and related cost models, whereas, for independent scalability, only a very few works explore the combination of horizontal and vertical scalability to support throughput requirements of the composite services. This can further consider the scalability constraints of different microservices (i.e., microservices with databases).
    \item Placement objectives of the current works lack emphasis on the following: security challenges related to microservice, utilising microservices for improvement of reliability and fault tolerance, dynamic microservice placement under federated Fog architectures, dynamic Fog-Cloud balance, mobility-aware placement, consideration for load uncertainty, etc.
\end{enumerate}

\section{Microservice Composition} \label{tax3}

\begin{figure}[t!]
    \includegraphics[width=0.6\linewidth]{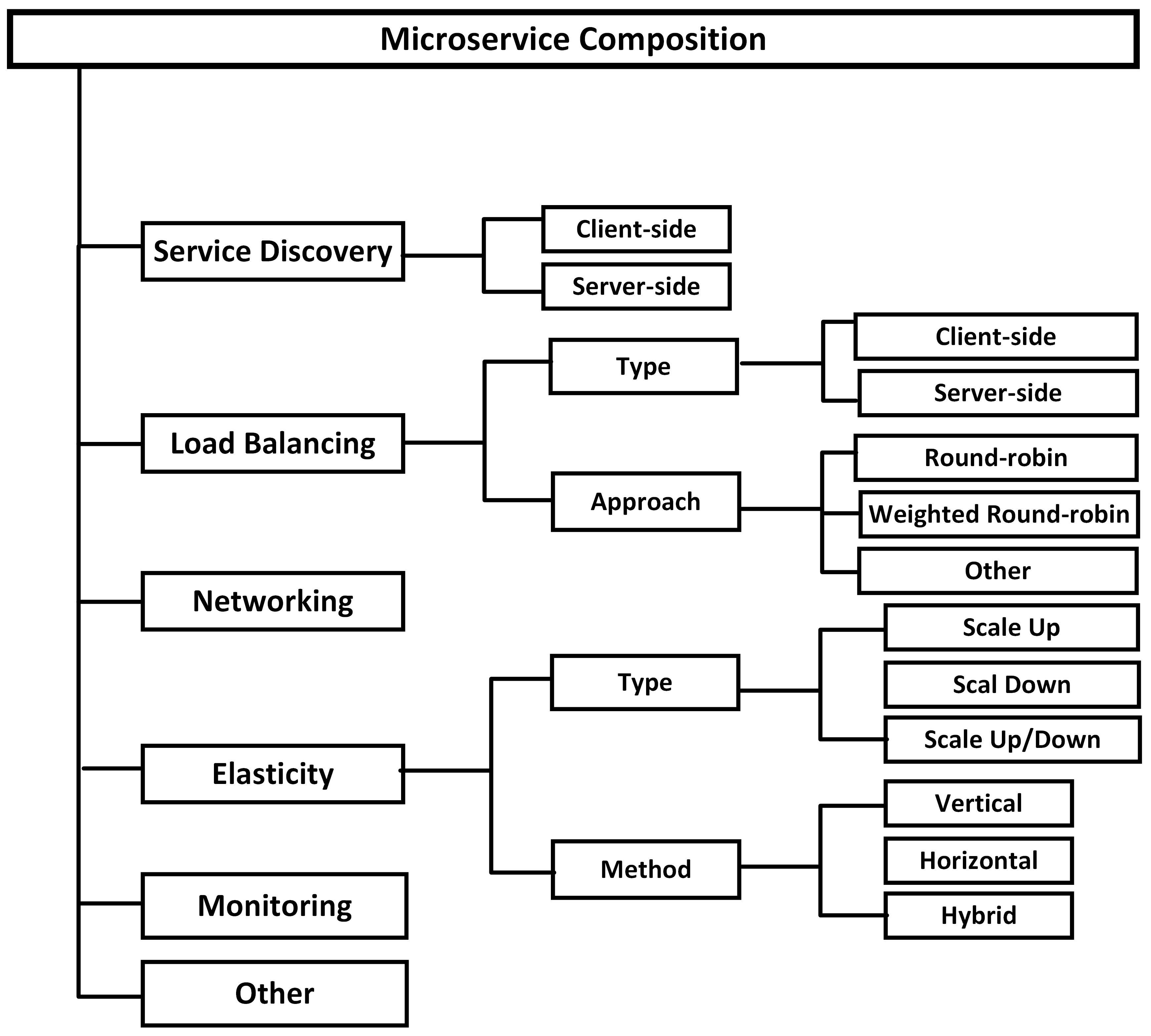}
    \caption{Taxonomy for microservice composition}
    \label{fig:compositionTaxonomy}
    \vspace*{-3.5mm}
\end{figure}

The complex interaction patterns among microservices, their ability to independently scale up/down to maintain performance and distributed deployment of microservice instances across networked devices are supported through microservice composition mechanisms. Fig. \ref{fig:compositionTaxonomy} presents the taxonomy for essential aspects of the successful composition of microservices within distributed environments. For this analysis, we use both Edge/Fog frameworks designed for microservice deployment \cite{fogatlas, yang2021kubehice} and conceptual frameworks/simulators used in formulating and solving placement problems \cite{deng2020optimal, he2021online} as demonstrated in Table \ref{table:tax3Works}. We analyse the main functions related to microservice composition in the sections below.

\begin{table*}[h]
	\caption{Analysis of existing literature based on the taxonomy for microservice composition}
	\label{table:tax3Works}
	\resizebox{\linewidth}{!}
	{\renewcommand{\arraystretch}{0.95}
	\begin{tabular}[htbp]{ |c|c|c|c|c|c|c|c|c|
	}
			\hline
			\multicolumn{1}{|c}{\textbf{Work}} & \multicolumn{1}{|c}{\textbf{Service}} &\multicolumn{2}{|c}{\textbf{Load Balancing}}
			&
			\multicolumn{1}{|c}{\textbf{Networking}}
			&\multicolumn{2}{|c}{\textbf{Elasticity}}
			&\multicolumn{1}{|c}{\textbf{Monitoring}}
			&\multicolumn{1}{|c|}{\textbf{Other}}
			\\
			\cline{3-4} \cline{6-7}
			 &  \textbf{Discovery} & 
			 \textbf{Type} 
			& \textbf{Approach} &  & Type  & Method & &
			\\
			\hline
			
			\cite{fogatlas} & S-S & S-S & Round Robin & \checkmark & U-D & H  &  \checkmark  & - \\
			& & & & (Kubernetes) & & & (Prometheus) & \\
			\hline
			
			\cite{pallewatta2019microservices} & C-S & C-S  & Weighted Round Robin & - & U & H+V & - & - \\
			\hline
			
			\cite{deng2020optimal} & NC & C-S & Round Robin & - & - & - & - & - \\
			\hline
			
			\cite{pallewatta2022qos} & NC & C-S & Weighted Round Robin & -& - & - & - & -  \\
			\hline
			
			\cite{armani2021cost} & S-S & S-S & Custom & \checkmark & - & - & \checkmark  & - \\
			& & & & (Flannel + Istio) && & (Prometheus) & \\
			\hline
			
			\cite{lv2022microservice} & S-S & S-S & Custom & \checkmark & U-D & H &  \checkmark & - \\
			& & &  & (Kubernetes + Istio) & && (Prometheus) & \\
			\hline
			
            \cite{huang2020ant} & ND & S-S & Custom & - & -&-	& - & \\
            \hline
            
            \cite{lei2020heuristic} & S-S & NC & - & - & -  &- &- & Fault-tolerance \\
            \hline
            
            \cite{he2021online} & ND & C-S & Round Robin & - & - & - & - & - \\
            \hline
            
            \cite{yang2021kubehice} & S-S & S-S & Weighted Round Robin & \checkmark & - & -& \checkmark  & - \\
            & & & & (Kubernetes + KubeEdge) & & & (Custom) &\\
            \hline
            
            \cite{mahmud2022ifogsim2} & C-S & C-S & Weighted Round Robin & \checkmark & U-D & H+V & - & - \\
            \hline
            
            \cite{de2019reactive} & S-S & S-S & ND & \checkmark & U-D & H & \checkmark  & Fault-tolerance \\
            & & & &  (MQTT Broker)& & & (Custom) & Security \\
            \hline
            
            \cite{buzachis2018towards} & S-S & S-S & ND & \checkmark & U-D & H & - & - \\
            & & & &  (Kubernetes + CNI)& & & & \\
            \hline
    \end{tabular}
	\renewcommand{\arraystretch}{1}}
    \raggedright\footnotesize{C-S:Client-side, S-S: Server-side, U: Up, D: Down, H: Horizontal, V: Vertical, CNI: Container Network Interface, ND: Not Defined, NC: Not Considered}
	\vspace*{-3.5mm}
\end{table*}

\subsection{Service Discovery} Dynamic placement algorithms proposed to handle microservice placement \cite{lv2022microservice, pallewatta2019microservices, yang2021kubehice} define service discovery mechanisms so that changes (i.e., scale up/down, failures) in microservice instances are made know to other microservices that interact with them. \cite{mahmud2022ifogsim2, pallewatta2019microservices} use a client-side service discovery pattern where each client microservice queries a dynamically updated service registry to determine the available service instances. Works such as \cite{lei2020heuristic, lv2022microservice, yang2021kubehice} use server-side service discovery where the requests are directed to a designated entity (i.e., a load balancer, a proxy, etc.) that is responsible for directing the requests towards available service instances. "FogAtlas"\cite{fogatlas} which is a Fog computing platform used by multiple works such as \cite{faticanti2019cutting, faticanti2020deployment} and other works such as \cite{yang2021kubehice, armani2021cost} use Kubernetes as the orchestrator along with its default proxy-based, server-side service discovery mechanisms. \cite{lei2020heuristic} implements server-side service discovery using an API gateway as the ingress node responsible for service discovery and service composition.

\subsection{Load Balancing}
\begin{enumerate}
   \item \textit{Type: } Current works model load balancing using two primary approaches: \textit{client-side load balancing} \cite{deng2020optimal, pallewatta2019microservices, he2021online} where each client is responsible for individually executing load balancing policies, thus enabling application dependent load balancing policies to be implemented, and \textit{server-side load balancing} \cite{armani2021cost, lv2022microservice, yang2021kubehice} where a dedicated load balancer sits between client and server microservices to handle load balancing. 
    \item \textit{Approach:} Integrating the effect of the load balancing mechanism and introducing novel load balancing policies to improve the performance of the services is vital in microservice application deployment. \cite{deng2020optimal, pallewatta2022qos} model the end-to-end service latency based on the existing load balancing policies such as Round Robin and Weighted Round Robin to determine the number of microservice instances to deploy. Meanwhile, some works introduce custom load balancing policies: \cite{armani2021cost} proposes a load balancing policy for a multi-region Fog architecture where the requests are directed based on the residual CPU of each Fog region, \cite{lv2022microservice} uses the variance of the resource occupancy rate of the edge nodes to determine where to direct the requests, \cite{huang2020ant} uses a meta-heuristic algorithm to place microservice replicas to identify flow paths by considering parameters such as service cost and service latency.
\end{enumerate}

\subsection{Networking} Distributed deployment of containerised microservices makes networking one of the integral functions of microservice compositions. This is further complicated by the federation of Fog and Cloud, which results in communication between multiple networking environments and technologies. \cite{fogatlas} use Kubernetes to handle the networking among microservices instances whereas \cite{lv2022microservice} integrates Istio\footnote{https://istio.io/} service mesh framework to handle inter-service communication. \cite{yang2021kubehice} integrate Kubernetes with KubeEdge for the edge network. To overcome the limitation Kubernetes network model and assign subnets per host, \cite{armani2021cost} use Flannel, a Container Network Interface (CNI) and Istio on top of Kubernetes orchestration. \cite{buzachis2018towards} explore and compare multiple CNIs, including Flannel, Weave, Calico and OVN. \cite{de2019reactive} uses MQTT Broker, an asynchronous messaging-based communication mechanism to transmit messages over the network among decoupled microservices.

\subsection{Elasticity} Elasticity indicates the ability of the microservices to be dynamically scaled up and down dynamically in a performance-aware manner. With the use of lightweight deployment technologies such as containers, microservices can be easily auto-scaled to improve the performance of the application while ensuring optimum resource utilisation. Out of the many works that consider horizontal scalability of microservices, the majority consider this during the initial placement of the application to make use of resource-constrained Fog devices but fail to use auto-scaling/elasticity under dynamic changes in the environment (i.e., load changes, failures etc.). Dynamic placement algorithms proposed in works such as \cite{lv2022microservice, mahmud2022ifogsim2, pallewatta2019microservices} consider elasticity. However, we can further analyse it based on the supported type of elasticity: scaling up, scaling down, and the method of scaling: horizontal, vertical. \cite{pallewatta2019microservices} considers both vertical and horizontal scaling but only considers scaling up as new user requests arrive. \cite{lv2022microservice} proposes a utilisation threshold-based policy to horizontally scale up/down microservices through continuous monitoring of the resources. Practical frameworks and simulators \cite{mahmud2022ifogsim2, fogatlas} provide the infrastructure required to auto-scale (both up and down, horizontal and vertical) microservices through customised policy implementations. 

\subsection{Monitoring } Monitoring is the collection of application and platform metrics in such a way that they can be used to detect failures, performance degraded states, etc. and act accordingly to maintain system performance. The highly dynamic nature of containerised microservices makes monitoring a challenging task, which has resulted in the development of open-source monitoring tools that can handle the large volume of moving parts in microservices-based application deployment. Hence, \cite{fogatlas, armani2021cost, lv2022microservice} integrate Prometheus\footnote{https://prometheus.io/} to their orchestration platform to monitor multiple platform metrics (i.e., request number, response times, resource consumption, network communications, etc.). Meanwhile, \cite{yang2021kubehice, de2019reactive} implement their own customised monitoring tools to monitor metrics across Edge-Cloud integration.

\subsection{Other} 
Other composition-related tasks include fault tolerance and security. \cite{lei2020heuristic} proposes a conceptual framework where API Gateway handles the responsibility of fault-tolerance by data recovery, service re-composition and re-submission of a failed request. \cite{de2019reactive} introduces three main components: health check component, circuit breaker and timeout component to identify and isolate failures to avoid cascading failures under MSA. \cite{de2019reactive} implements centralised security components to manage authorisation and authentication required for microservice access.

\subsection{Research Gaps}
Based on the analysis of existing works presented in Table \ref{table:tax3Works}, we identified the following gaps:
\begin{enumerate}
    \item Current works demonstrate less emphasis on load-balancing policies and their effect on the formulation of the placement problem. 
    \item There's scope for improvement in elasticity (both horizontal and vertical scaling), fault tolerance, and microservice security through proper monitoring of the deployed application at the application and platform level.
    \item Effect of orchestration components ( i.e., service registry, API gateways, load balancers) on the application performance needs to be analysed and demonstrated in terms of handling their possible failures, delay overheads, etc.
\end{enumerate}

\section{Performance Evaluation} \label{tax4}

\begin{figure}[h]
    \includegraphics[width=0.55\linewidth]{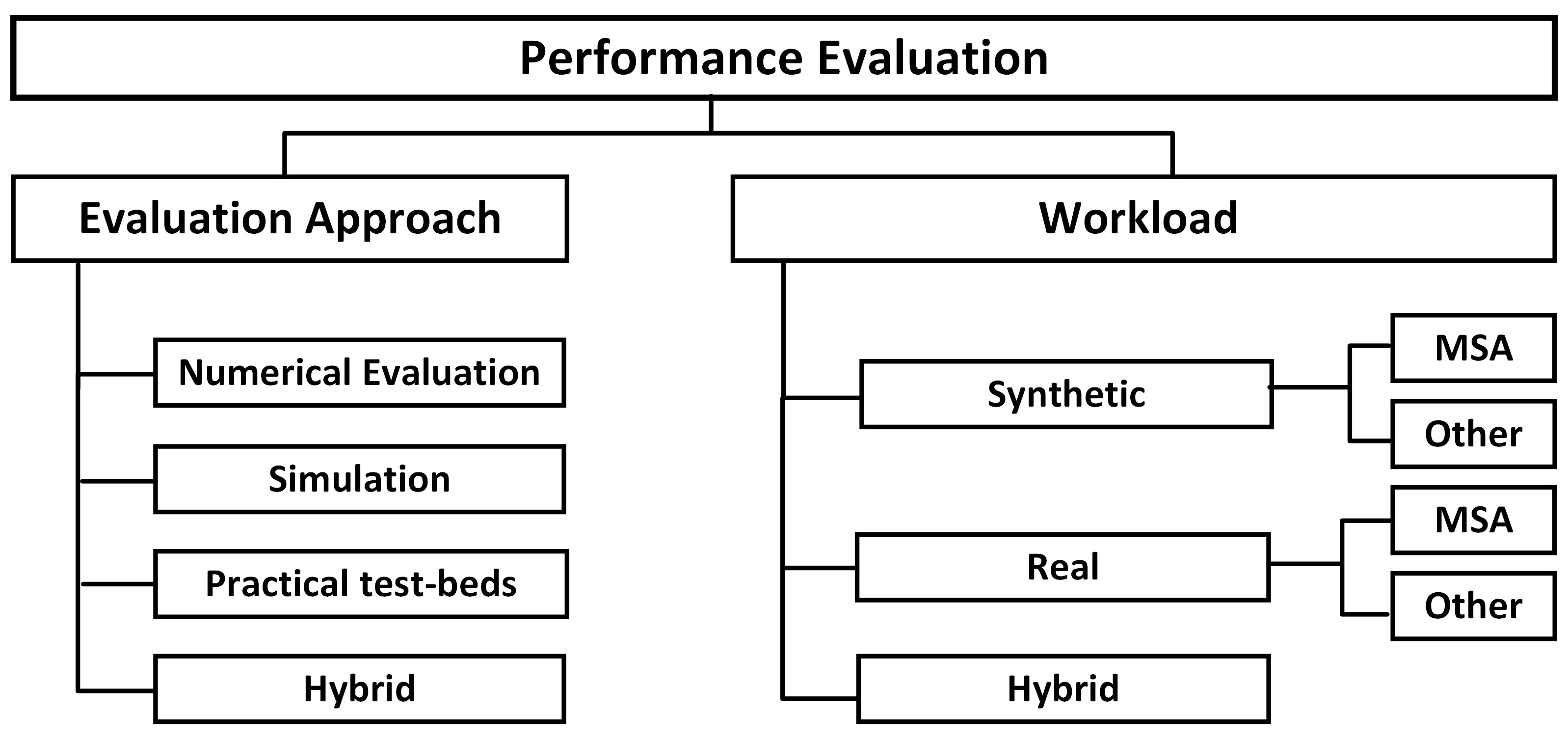}
    \caption{Taxonomy for performance evaluation of the placement policy}
    \label{fig:evalTaxonomy}
    \vspace*{-3.5mm}
\end{figure}

\begin{table*}[h]
	\caption{Analysis of existing literature based on the taxonomy for performance evaluation }
	\label{table:tax4Works}
	\resizebox{\linewidth}{!}
	{\renewcommand{\arraystretch}{1}
	\begin{tabular}[htbp]{ |c|c|c|c|c|c|c|||c|c|c|c|c|c|c|
	}
			\hline
			\multicolumn{1}{|c}{\textbf{Work}} &
			\multicolumn{3}{|c}{\textbf{Evaluation Approach}} &  \multicolumn{3}{|c|||}{\textbf{Workload}}
			& 
			 
			\multicolumn{1}{c}{\textbf{Work}} &
			\multicolumn{3}{|c}{\textbf{Evaluation Approach}} &  \multicolumn{3}{|c|}{\textbf{Workload}}
			\\
			\cline{2-7} \cline{9-14}
			 & \textbf{Numerical} & \textbf{Simulation}  & \textbf{Practical} 
			& \textbf{Synthetic} & \textbf{Real}  & \textbf{Hybrid} & & \textbf{Numerical} & \textbf{Simulation}  & \textbf{Practical} 
			& \textbf{Synthetic} & \textbf{Real}  & \textbf{Hybrid}
			\\
			\hline
			\cite{lera2018availability} & & \checkmark (YAFS) & & \checkmark (Other) & &  &
			\cite{faticanti2019cutting} & \checkmark &- & \checkmark (FogAtlas) & \checkmark (MSA) &- & - \\
			\hline
			
			\cite{pallewatta2019microservices} & - & \checkmark (iFogSim) & - & \checkmark (MSA) & -& -  &
			\cite{guerrero2019evaluation} & \checkmark  & - & - & - & -& \checkmark  \\
			\hline
			
			\cite{faticanti2020throughput} & \checkmark & - & -& \checkmark (Other) & -  & - & 
			\cite{zhao2020distributed} & - & \checkmark (\textbf{$C$}-MATLAB) & - & \checkmark (MSA) & - & - 
			\\
			\hline
			
			\cite{deng2020optimal} & - & \checkmark  & - & \checkmark (MSA) & -& - &
			\cite{wang2019delay} & - & \checkmark (\textbf{$C$}-MATLAB) & - & \checkmark (MSA) & -& - \\
			& & (MATLAB + CloudSim) & & & & & & & & & & & \\
			\hline
			
			\cite{pallewatta2022qos} & - & \checkmark (iFogSim) & - & \checkmark (MSA) & -& -  &
			\cite{armani2021cost} & - & \checkmark (\textbf{$C$}-Python) & \checkmark & - & -& \checkmark \\
			\hline
			
			\cite{fang2020iot} & - & \checkmark (iFogSim) & - &  \checkmark (MSA) & -& - &
			\cite{samanta2020dyme} & - & \checkmark (ND) & - &  \checkmark (MSA) & - & - \\
			\hline
			
            \cite{lv2022microservice} & - & - & \checkmark &  - & \checkmark (MSA) & - &
            \cite{abdullah2020predictive} & - & - & \checkmark & - & - & \checkmark \\
            \hline
    
            \cite{xu2021adaptive} & - & \checkmark (\textbf{$C$}-Java) & - & - & \checkmark (Other) &  - &
            \cite{huang2020ant} & - & \checkmark (ND) & - & \checkmark (MSA) & -& - \\
            \hline
            
            \cite{herrera2021optimal} & \checkmark & - & - & \checkmark (MSA) & - & - & 
            \cite{he2021online} & \checkmark & - & - & \checkmark (Other) & - & - \\
            \hline
            
            \cite{lei2020heuristic} & - & \checkmark (\textbf{$C$}-Python) & - & - & \checkmark (MSA) & - & 
            \cite{alencar2021dynamic} & - & \checkmark (NS3) & - &  \checkmark (MSA) & - & - \\
            \hline
            
            \cite{faticanti2020deployment} & - &  \checkmark (\textbf{$C$}-Python) & - & \checkmark (Other) & - & - & \cite{filip2018microservices} & -& \checkmark (Cloudsim) & - & \checkmark (MSA) &  - & - \\
            \hline
            
            \cite{guo2022joint} & \checkmark & - & - & \checkmark (MSA) &  - & - & \cite{rodriguez2021improvement} & - & \checkmark (iFogSim) & - & - & \checkmark (MSA) & - \\
            \hline
            
            \cite{fu2021qos} & - & - & \checkmark & - & \checkmark (MSA) & - & 
            \cite{yang2021kubehice} & - & - & \checkmark & - & \checkmark (MSA) & - \\
            
            \hline
            \cite{zhang2022astraea} & - & - & \checkmark (Astraea) & - & \checkmark (MSA) & - &
            \cite{guerrero2019lightweight} & - & \checkmark (iFogSim) & - & - & - & \checkmark \\
            
            \hline
            \cite{paul2020crew} & - & \checkmark (iFogSim)& - & - & - & \checkmark &
            \cite{xu2020service} & - & \checkmark($C$-C++) & - & \checkmark (MSA) & - & - \\
            \hline
	\end{tabular}}
	\vspace*{-3.5mm}
\end{table*}

Accurate policy evaluation is one of the vital steps in designing novel placement algorithms for fast-evolving fields such as IoT and Fog computing. To this end, we propose the final taxonomy by analysing the crucial aspects of the evaluation phase, as shown in Fig. \ref{fig:evalTaxonomy}. Afterwards, current works are mapped to the taxonomy to identify gaps and possible improvements (see Table \ref{table:tax4Works}).

\subsection{Evaluation Approach}
Evaluation approaches can be categorised as Numerical Evaluations, Simulations, use of Practical test-beds and Hybrid approaches consisting of combinations of the above methods.

\textit{Numerical Evaluations}: In numerical experiments, the algorithm is evaluated by numerically calculating specific metrics that provide insights on the fitness of the resultant placement proposed by the algorithm \cite{faticanti2019cutting, faticanti2020throughput, he2021online, herrera2021optimal} or/and evaluating performance metrics of the algorithm such as execution time, computation complexity and convergence \cite{guerrero2019evaluation, he2021online}. \cite{faticanti2019cutting, faticanti2020throughput} use the Gurobi mathematical optimisation solver to obtain the optimum solution to the formulated MINLP and compare the solution obtained from their proposed heuristic placement algorithm by calculating metrics such as the number of placed applications, network link usage, etc. numerically. \cite{guerrero2019evaluation} evaluates their approach based on the performance metrics (i.e., the fitness of the best solution, Pareto solution spread, execution time, etc.) of their proposed multi-objective evolutionary algorithm, whereas \cite{he2021online} analyses the execution time of the algorithm under different experimental settings to evaluate the scalability of the algorithm, thus deriving the system size the algorithm can handle. 

\textit{Simulations}: For the evaluation of microservices-based application placement in Edge/Fog computing environments, both open-source simulators (i.e., iFogSim \cite{fang2020iot, guerrero2019lightweight, pallewatta2019microservices, pallewatta2022qos}, YAFS \cite{lera2018availability}, CloudSim \cite{filip2018microservices, deng2020optimal}) and custom simulators \cite{zhao2020distributed, wang2019delay, armani2021cost} are used by the current works. Simulators such as iFogSim \cite{mahmud2022ifogsim2} and YAFS \cite{lera2019yafs} provide in-built microservice-related features such as distributed application modelling, service discovery and load balancing, thus enabling the users to simply implement their placement policy within the simulator or implement the algorithm separately (i.e., MATLAB \cite{deng2020optimal}, IBM CPLEX \cite{pallewatta2022qos}) and input the resultant placement to the simulator for evaluations. 

\textit{Practical}: To evaluate placement algorithms using practical frameworks, \cite{faticanti2019cutting} uses "FogAtlas", an open source framework for microservices deployment and orchestration in Fog environments, \cite{zhang2022astraea} implements a  framework called "Astraea" for the management of GPU microservices, whereas other works such as \cite{abdullah2020predictive, armani2021cost, lv2022microservice, yang2021kubehice} implement customised test-beds with functionalities relevant to the placement algorithms. They use popular container-orchestration platforms such as Docker swarm \cite{abdullah2020predictive}, Kubernetes \cite{armani2021cost, lv2022microservice}, KubeEdge \cite{yang2021kubehice} in their implementations.  

\textit{Hybrid}: Some of the works use multiple evaluation approaches to analyse and evaluate the proposed placement policies from multiple perspectives.  \cite{armani2021cost} use both Simulations and practical test beds for evaluation. Simulations carry out large-scale experiments, whereas test beds further verify the results of the simulations by carrying out a selected set of experiments. \cite{pallewatta2022qos} use a combination of numerical evaluations and simulations where numerical evaluations are used to improve and fine-tune the meta-heuristic placement algorithm, whereas the simulation evaluates the resultant placements.

\vspace*{-2.5mm}
\subsection{Workload}

For the analysis of the workload, we consider the nature of the application placement requests used to evaluate the placement policy. We can categorise them as Synthetic, Real, or a combination of both, denoted as Hybrid.

\textit{Synthetic}: Synthetic workloads are created either by mimicking specific microservices-based applications \cite{filip2018microservices, faticanti2019cutting, pallewatta2019microservices} (categorised in the taxonomy as MSA) or using generic application models such as DGs or DAGs \cite{lera2018availability,faticanti2020deployment, he2021online, faticanti2020throughput} (categorised in the taxonomy as Other) to generate a workload consisting of multiple applications with heterogeneous resource and QoS requirements. \cite{faticanti2019cutting} models the applications following a microservices-based IoT application for face recognition consisting of two chained microservices. \cite{pallewatta2019microservices} models a smart-healthcare application and creates a synthetic workload based on the modelled application. \cite{filip2018microservices} models smart city and forest surveillance applications. Meanwhile, \cite{faticanti2020throughput, faticanti2020deployment, he2021online} generate random synthetic DAGs as microservices-based applications, whereas \cite{lera2018availability} uses Growing Network(GN) graph structure where graphs are created by adding nodes one at a time to existing nodes to develop microservices-based applications following Directed Graphs as the interaction pattern.

\textit{Real}: Real workloads include the use of already implemented microservices-based applications (categorised in the taxonomy as MSA) or adapting performance traces of applications that follow other application models (categorised in the taxonomy as Other). \cite{lv2022microservice} use Bookinfo \footnote{https://istio.io/latest/docs/examples/bookinfo/}, an online book store application following MSA along with the hotel reservation booking application from DeathStartBench \footnote{https://github.com/delimitrou/DeathStarBench/tree/master/hotelReservation} \cite{gan2019open}, which is a benchmark application suite following MSA. \cite{fu2021qos} also uses benchmark applications from DeathStartBench along with the benchmark application provided in \cite{zhou2018fault}. \cite{zhang2022astraea} uses AI-based GPU microservices available in AIbench \footnote{https://www.benchcouncil.org/aibench/index.html} to create the workload. \cite{lei2020heuristic} uses the curated data set available in \cite{rahman2019curated} which consists of 20 open-source projects based on MSA. In contrast to the above examples, \cite{xu2021adaptive} uses traces from Google Cluster 2019 \cite{tirmazi2020borg}. These traces provide data on task requests (i.e., CPU. memory, deadline, etc.), and as \cite{xu2021adaptive} models microservices as independent components that do not interact with other microservices, the said data set is easily adapted by this work for evaluations.

\textit{Hybrid}: \cite{guerrero2019evaluation, guerrero2019lightweight} use a microservices-based e-commerce application known as Sock Shop \footnote{https://microservices-demo.github.io/} provided under Apache License 2.0, along with two other synthetic application models (an online EEG tractor beam game and intelligent surveillance application) to create the workload. \cite{armani2021cost} creates a synthetic workload for simulation-based studies and implements a microservices-based application name "Paper Miner" designed for mining research papers and deploys it on a real-world platform to evaluate the placement algorithm.

\subsection{Research Gaps}
Based on the analysis of existing works presented in Table \ref{table:tax4Works}, we identified the following gaps related to
the evaluation of placement algorithms developed for microservices-based application placement within Fog environments.
\begin{enumerate}
    \item Lack of use in practical test beds is one of the prominent drawbacks of currently used evaluation approaches. The majority of the available works use numerical evaluations or simulations to evaluate the performance of their placement policies but fail to validate them on real test beds. Thus, overheads related to orchestration tasks (i.e., service discovery, load balancing, auto-scaling etc.), failure characteristics, resource contention among microservices, etc., are not accurately captured. Moreover, the suitability of the Edge/Fog devices to act as the placement engine that runs the algorithms is not evaluated in practical settings.
    
    \item As Edge/Fog computing paradigms are still relatively new and yet to be adopted by the service providers, simulators play a significant role in evaluating placement policies. However, this requires a standard open-source simulator for use among the research community and continuous improvements through collaboration. As microservices-based application placement in Fog environments is still in its infancy, we see the increased use of custom simulators due to the lack of open-source simulators that capture all related aspects of microservice orchestration.
    
    \item Lack of real-world traces or actual implementations of applications for the deployment within test beds is another significant gap in current research. The existing benchmark applications do not include IoT applications, making it harder to capture their characteristics accurately. 
\end{enumerate}

\section{Future Research Directions} \label{futureD}
Based on the research gaps identified in previous sections, we propose future directions for microservices-based IoT application placement in Fog environments.

\hfill

\textbf{\textit{Dynamic placement algorithms -}} To maintain application QoS under the dynamic nature of the Fog infrastructure and fluctuating workloads, application management algorithms must adapt and make decisions accordingly. Online placement algorithms that carry out continuous re-evaluation of the application placements can achieve this. Algorithms can exploit the independently deployable nature of the microservices, which adds dynamic behaviour to them through auto-scaling, migration, proactive redundant placements, Fog-Cloud balanced usage, etc. To this end, the placement techniques can benefit from AI-based techniques, such as evolutionary algorithms and ML techniques, such as Reinforcement Learning, that can adapt to dynamic environmental changes. 

\hfill

\textbf{\textit{Placement within federated multi-fog multi-cloud environments -}} Federation among Fog resources provided by multiple Fog infrastructure providers (multi-fog) and multi-cloud environments is emerging as an approach better suited for utilising geo-distributed and resource-constrained Fog computing resources to meet the non-functional requirements of IoT applications. The loosely coupled nature of the microservices allows them to span across such environments while maintaining seamless connectivity among them. However, for such scenarios, placement policies need to consider costs, resource availability, security, composition-related overheads and limitations based on the infrastructure provider and location of the Fog resources. Moreover, placement policies must evolve towards distributed placement approaches to handle placement within multi-fog environments.  

\hfill

\textbf{\textit{Microservice orchestration platforms for Fog -}} For the evaluation of scheduling policies within Cloud environments, commercial platforms such as AWS, Google Cloud, and Microsoft Azure are available. As Fog computing is still in its early stages of industrial adaptation, current research uses small, custom-built test beds. However, they lack support for large-scale experiments, thus failing to capture important aspects related to MSA, such as distributed, location-aware deployments, load balancing, reliability, security and interoperability of services within the large-scale IoT ecosystem. Moreover, they should reflect novel technologies (i.e., container orchestration, service mesh, monitoring tools, overlay networks, etc.). Hence, scalable and extensible container orchestration platforms for Fog environments should be implemented for research purposes.

\hfill

\textbf{\textit{IoT workloads/ benchmarks related to MSA - }} Lack of microservices-based IoT workload traces from large-scale deployments and benchmark IoT applications that follow MSA hinder the evaluation of placement policies considerably. Enterprise workload traces of IoT applications can be used to derive accurate data related to request volumes, patterns, diversity in usage of services, etc. Collecting such data over a long time within large-scale IoT deployments and making them accessible to the research community is significant for accurately evaluating placement algorithms.

\hfill

\textbf{\textit{Security-aware placement -}} Data privacy is one of the main concerns of data-driven IoT applications. Distributed deployment of microservices across Fog-Cloud, along with the vulnerability of open microservice interfaces, poses a considerable security threat to sensitive data transmission and processing. The independently deployable nature of microservices enables the migration of microservices easily across federated Fog environments and between Fog and the Cloud. However, placement algorithms have to incorporate data privacy and security threats related to such migrations in making deployment decisions.

\hfill

\textbf{\textit{Resource contention handling -}} Due to the concurrent execution of multiple containers within the same device, resource limitation in the devices and complex interaction patterns of the microservices, resource contention among microservices can affect application performance negatively. The development of intelligent algorithms that can proactively identify resource contentions during microservice placement, dynamically auto-tune container parameters or migrate containers across the Fog-Cloud continuum has the potential to mitigate this challenge. 

\hfill

\textbf{\textit{Observability and monitoring driven maintenance -}} Deployment of microservices-based IoT applications creates a distributed system of many microservice instances that can be dynamically created and destroyed. Observability and monitoring can be used to detect performance anomalies within such systems. This requires distributed tracing, monitoring and analysis of the system at both application and platform levels, which would create massive amounts of data of different metrics, logs and traces. This poses a big data analysis challenge where data mining and artificial intelligence models can be integrated with the placement policy to make performance-aware decisions in handling performance anomalies and failures within the system. 

\hfill

\textbf{\textit{Placement within NFV-enabled networks -}} Network Function Virtualisation (NFV) has become a key enabler for 5G and 6G mobile networks. NFV virtualises network functions (i.e., firewalls, routers, load balancers, etc.) to provide flexible management and orchestration of network resources, thus supporting IoT applications with large, fluctuating traffic volumes to meet expected QoS levels. To this end, virtualised network functions are developed as containerised microservices that become part of the composite application services (known as service function chains). This requires microservices-based application placement and request routing to be solved in the context of service function chains, where the dynamic placement of containerised network functions becomes a significant part of microservices-based IoT application placement to improve dynamic traffic routing, security, etc., to satisfy the non-functional requirements of the application services.

\hfill

\textbf{\textit{Fault tolerant placement and management of microservices}}  - MSA avoids a single point of failure by decomposing applications into loosely coupled microservices. While this improves fault tolerance, it creates more points of failure and also cascading failures due to interacting microservices. Root-cause tracing, predictive redundant placements, awareness of microservice level stability patterns (i.e., circuit breaking, timeouts, retries, etc.) and their effect can be used to generate more resilient placement, migration and request load balancing approaches to improve fault-tolerance of IoT application services. 

\hfill

\textbf{\textit{Availability assurance under continuous integration and delivery -}} One of the main reasons for the rising popularity of MSA  is its ability to support rapid development and deployment cycles to keep up with the fast-evolving IoT domain. However, this requires smooth integration and deployment of novel updates and changes while minimising service downtime. MSA handles this through multiple deployment strategies such as Canary deployments, Rolling deployments, Blue-Green deployment and A/B testing. To support continuous integration and delivery requirements in an availability-aware manner, continuous placement policies need to be developed, including combinations of multiple deployment strategies supported by MSA.

\section{Summary}\label{summary}
In this paper, we discussed IoT applications designed and developed using MSA and their placement within Fog computing environments. We conducted a comprehensive background study and identified four critical aspects of microservices-based application placement: modelling of MSA, placement policy creation, microservice composition and performance evaluation. We proposed taxonomies for each of the aspects, highlighting features related to MSA. Moreover, we analysed and discussed the current literature under each taxonomy and identified research gaps. Finally, we compiled future research directions to improve the placement of microservices-based IoT applications in Fog computing environments.

\bibliographystyle{ACM-Reference-Format}
\bibliography{main}
\newpage
\appendix
\section{Background}
This section presents a comprehensive background on MSA, IoT applications and different computing paradigms like Edge computing, Fog computing, Mobile Edge Computing and Osmotic Computing to highlight their characteristics. Moreover, this section compares microservices-based application placement and the traditional DAG-based workflow scheduling to highlight the differences between the two problems.

\subsection{Comparison between Microservice and Monolithic Architecture for Applications}\label{microservices}

\begin{figure}[h]
  \begin{subfigure}{0.49\linewidth}
    \includegraphics[width=\linewidth]{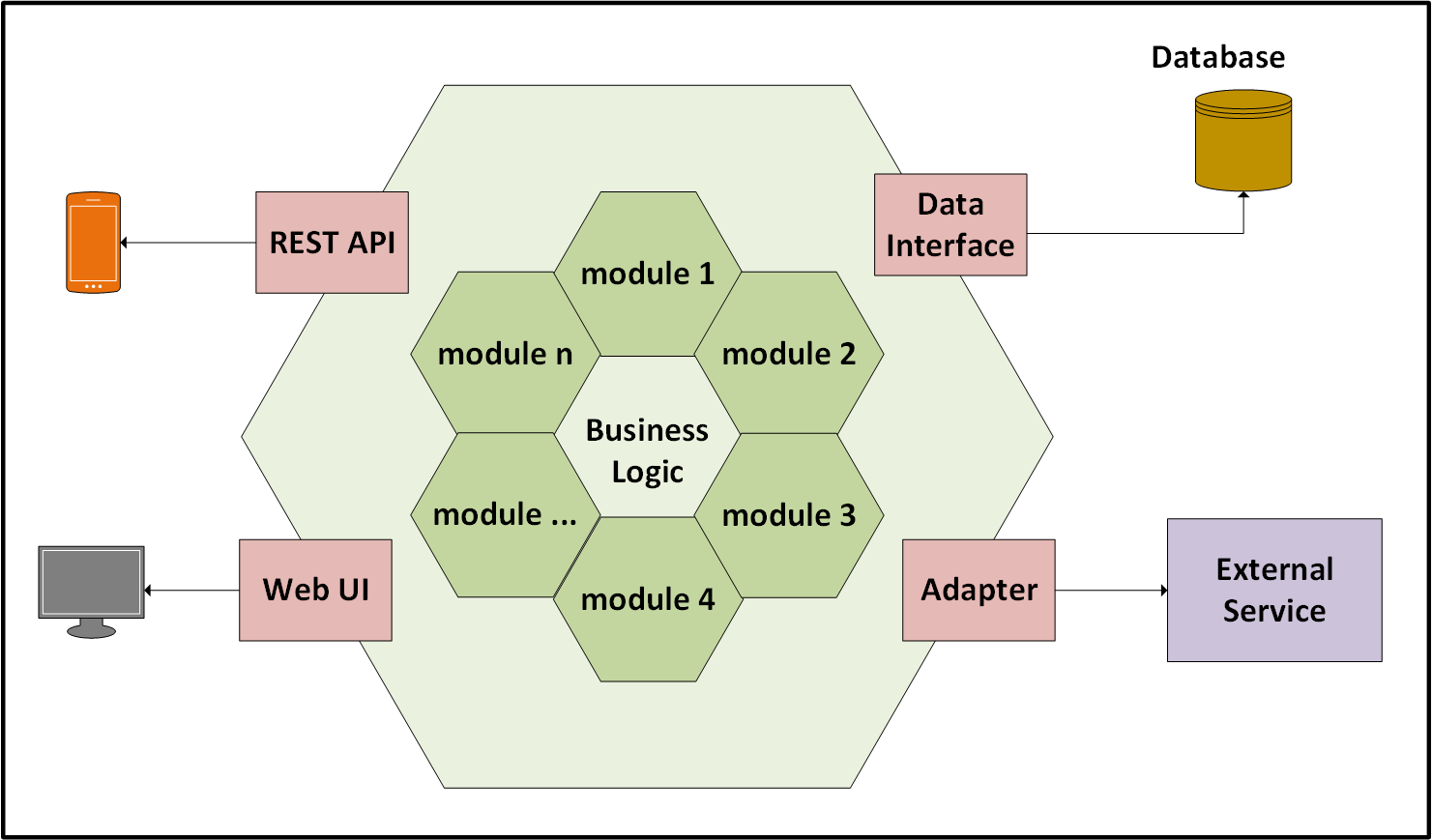}
    \caption{Monolithic application architecture}
    \label{fig:monlith}
  \end{subfigure}
  \hfill
  \begin{subfigure}{0.49\linewidth}
    \includegraphics[width=\linewidth]{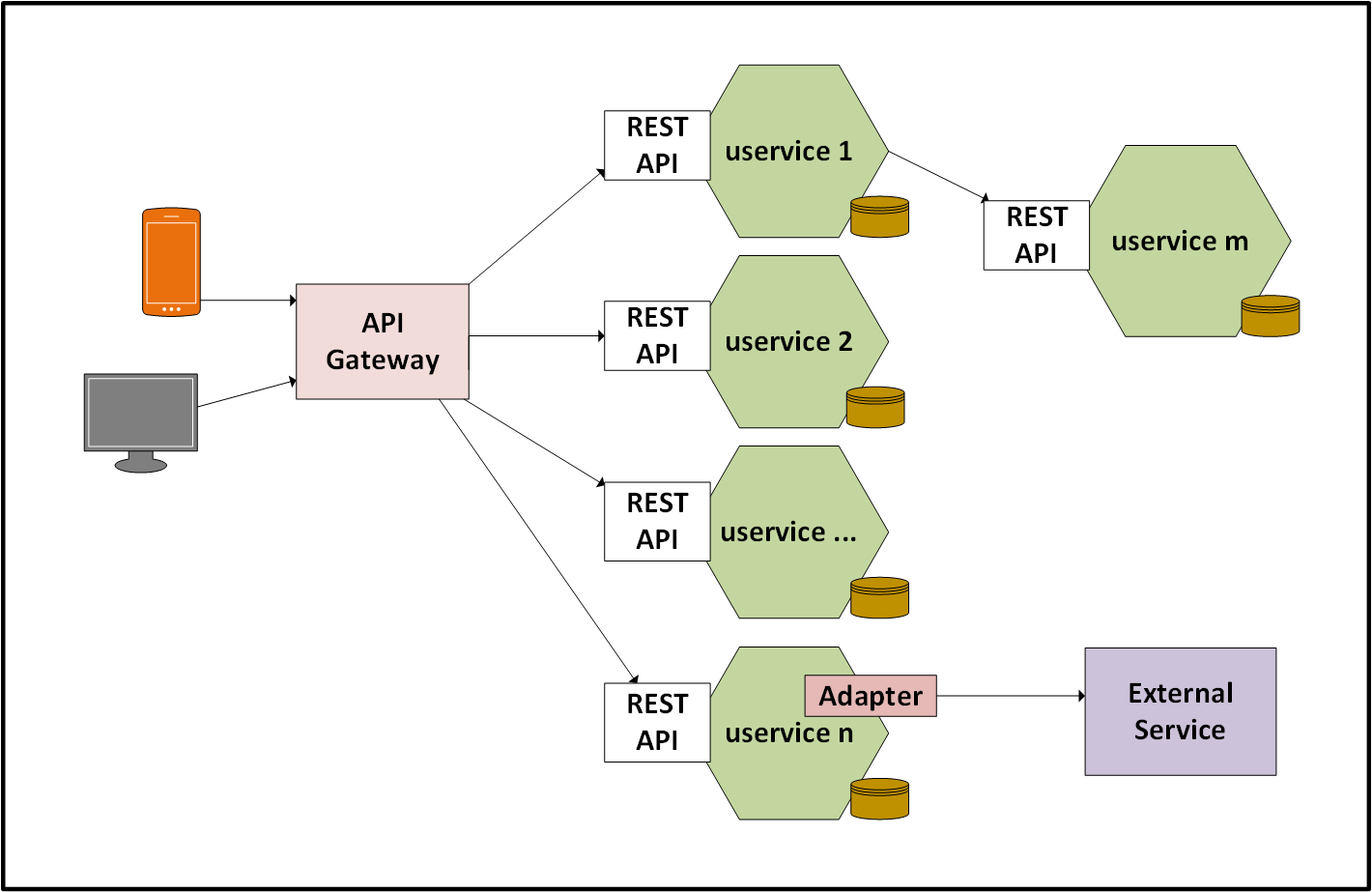}
    \caption{Microservice application architecture (MSA)}
    \label{fig:microservice}
  \end{subfigure}
  \caption{Application Architecture (Monolithic vs Microservices)}
  \label{fig:monolith vs microservice}
  \vspace*{-3.5mm}
\end{figure}

An enterprise application usually consists of a server-side application that implements its domain-specific business logic, client-side user interfaces supporting different clients such as desktop browsers and/or mobile browsers, and a database to persist the data. Moreover, the application may connect with other third-party applications (through web services or message brokers) and expose APIs for third parties to consume. The design and development of such applications follow different architecture patterns depending on the complexity of the business domain. Fig. \ref{fig:monolith vs microservice} presents a general representation of an application developed using  Monolithic and Microservice architecture.

In monolithic architecture, the server-side application is a single logical executable that handles HTTP requests from the client side, performs logical operations, communicates with the database and populates HTML views to be displayed on the client side. The server-side application can be designed and developed either as a single process or a modular monolith where modules invoke each other through method/function calls at the programming language level \cite{newman2019monolith}. Using monolithic architecture is advantageous during the early stages of an application or if the application is quite simple with only a few core functionalities. It provides a better approach to defining requirements and building, testing and deploying a working product quickly and efficiently. 

But as the applications grow, they tend to outgrow the monolithic architecture \cite{richardson2018microservices}. As all modules of the application are combined into a single unit, the coupling among them increases, and the application becomes too complex to understand, which makes it hard to make changes and improvements to the application. As any modification to the application affects the entire application, end-to-end testing becomes complex, resulting in high deployment delays. Moreover, monolithic architecture lacks flexibility as it requires a single development stack which hinders the adaptation of new technologies or heterogeneous technologies depending on the functionality of each module. Monoliths also lack reliability because a fault in one place results in the failure of the entire application. Moreover, different modules inside a monolith may have different resource requirements, which makes the application challenging to scale. Thus, monolithic architecture lacks support for agile development and limits the rapid growth of applications.

MSA aids in overcoming the limitations of the monolithic architecture within rapidly growing software ecosystems. Martin Fowler defines MSA as "an approach to developing a single application as a suite of small services, each running as independent processes and communicating with lightweight mechanisms, often an HTTP resource API" \cite{websiteFowler}. Fig. \ref{fig:microservice} depicts a microservice application developed using the API design pattern where all services are exposed to clients through an API gateway that directs client requests toward microservices. MSA is an evolution of Service Oriented Architecture (SOA), where microservices are fine-grained compared to services in SOA. It uses dumb pipes for the communication among them, unlike in the case of SOA, where services communicate using smart pipes containing business and message processing logic \cite{websiteFowler}. 

Thus, microservices are independently deployable and scalable units that are designed around business logic adhering to the single responsibility principle and contain their own database for persisting data \cite{joseph2019straddling}. Moreover, microservices are loosely connected components that can be easily integrated to create complex applications. From the software design and development perspective, this allows small teams to work on each service with minimum dependencies among the teams, thus increasing development velocity by supporting independent development and testing of microservices. Due to fine-grained modularity, each microservice can be deployed on hardware that best matches its resource requirements (i.e., CPU-intensive, memory-intensive, I/O-intensive, etc.) and can be deployed and scaled independently according to the load of each service. Unlike monolithic architecture, MSA has better fault isolation, as a fault in a certain service only affects that service. Furthermore, the independent scalability of microservices improves redundant deployment to achieve fault tolerance. As services are loosely coupled, new technologies that best suit the microservice can be easily adapted. This mitigates the need to stick to a single technology stack and provides the flexibility to evolve with technologies. Table \ref{table:Comparison1} summarises and compares the characteristics of the two architectures.

\begin{table}[t!]
	\caption{Comparison between Monolithic and Microservice architecture}
	\label{table:Comparison1}
	\centering
    \footnotesize
	\resizebox{\linewidth}{!}
	{\begin{tabular}{p{0.4\textwidth}p{0.6\textwidth}}
			\hline
			\textbf{Monolithic Architecture} & \textbf{Microservice Architecture} \\
			\hline
			Server-side application is a single logical executable & Server-side consists of independently deployable, multiple microservices\\
			Modules communicate through language level method invocations & Inter-process communication through REST APIs or lightweight messaging  \\
			Low flexibility & Flexible - different programming languages and technologies can be used based on microservice requirements \\
			Less support for scaling & Highly scalable (independently deployable and scalable microservices) \\
			Longer time from development to deployment & Supports rapid and agile development and deployment \\
			Unreliable due to single point of failure & Higher resilience to failure due to failure isolation and redundancy support\\
			\hline
	\end{tabular}}
	\vspace*{-6.0mm}
\end{table}

MSA enables continuous development and deployment of rapidly evolving complex applications but introduces novel challenges in design, development, deployment and orchestration, described in detail below:
\begin{itemize}
    \item \textit{\textbf{Microservice design}}: From a software design perspective, defining microservice boundaries can be identified as one of the main challenges of MSA. The level of granularity directly affects the performance of the applications. Dividing the application logic into too many microservices increases the number of remote calls made among microservices, thus increasing the latency of the operations. In contrast, less granularity creates macro modules with lower cohesion and higher coupling. Hence, an optimum level of granularity needs to be achieved by using the domain-driven approach for the separation of business concerns so that extensibility and adaptability are supported as the application requirements evolve \cite{hassan2020microservice, shadija2017microservices}. Moreover, it results in designing microservices as independently deployable and scalable components.
    
    \item \textit{\textbf{Application Placement}}: Granularity creates complicated dependencies among microservices and complex composition patterns, which results in composite services. Openness in microservice design allows reusability of microservices between multiple composite services with heterogeneous QoS requirements \cite{niu2018load}, conditional branching, which results in alternative data flows within composite services \cite{oberhauser2016microflows}, and use of third-party microservices \cite{al2018enhancing}. These characteristics need to be considered during the application placement to improve performance (i.e., QoS, resilience, scalability, security, etc.) while satisfying competing performance requirements of the services. Moreover, characteristics of the microservices, such as their independently deployable and scalable nature, have the potential to improve performance if appropriately utilised by the placement algorithms. Thus, novel placement algorithms are required to handle the above complexities efficiently while using the benefits provided by the architecture \cite{pallewatta2019microservices, villari2016osmotic}.
    
    \item \textit{\textbf{Application management}}: The increase in the number of moving components per application makes management more complex. At the same time, due to the elasticity of microservices, service instances can be dynamically scaled up and down with the changing workload or system failures. This requires support for automated deployment using orchestration platforms. To overcome the complexities introduced by the number of microservices and their dependencies, proper monitoring is crucial to provide insights (i.e., the health of the microservices/APIs, isolation of failures, demand variations etc.) to make management decisions.
    
    \item \textit{\textbf{Microservices communication}}: With the development of applications as a collection of interconnected modules, proper communication among microservices must be established using different communication patterns. Suitable communication patterns are selected by analysing the nature of the interactions among microservices \cite{shadija2017microservices}. Such patterns include Synchronous communication (i.e., request-response methods like HTTP-REST),  Publish-subscribe and Asynchronous communication using message queues (RabbitMQ, Apache Kafka, and Apache ActiveMQ \cite{de2019performance}).
    
    \item \textit{\textbf{Service discovery}}: Within a microservices application, microservices communicate with each other to perform tasks. This requires client microservices to be aware of other available microservices that can be invoked to perform specific tasks. With the distributed nature of the microservices and the dynamism of microservices placement  (auto-scaling,  migration,  service failures),  service discovery has become one of the main challenges in using MSA. Dynamic service discovery involves checking the status of existing microservices and detecting and registering new microservices as they become available in real-time, with minimum overhead.
    
    \item \textit{\textbf{Load balancing}}: MSA supports horizontal scalability or a combination of horizontal and vertical scalability of the microservices. As a  result, to ensure performance,  load balancing is required to direct requests among horizontally scaled microservices. Efficient load balancing algorithms have to be designed to minimise the networking and computation overhead and ensure microservice availability through load analysis, link analysis, and forecasting to adapt to dynamic scenarios \cite{wang2021research, niu2018load}. Thus, load balancing is tightly coupled to service discovery and monitoring of the microservices-based systems.
    
    \item \textit{\textbf{Fault tolerance}}: Even though MSA supports higher fault isolation compared to monolith architecture, it still requires fault tolerance methods to handle failures. Having a large number of components means there are more points of failure. Moreover, communication failures among components become a potential cause as well. Fault tolerance mechanisms should be able to tackle such failures and avoid cascading failures as well. To achieve this, MSA aims to handle failures at both the development and deployment phases. Microservice applications integrate methods such as circuit breakers, retries, timeouts and deadlines to the microservice design to minimise the effect of failures, whereas redundant placement, checkpoint and restore, migration, etc., are used at deployment to improve the resilience and availability of the application under failures. 
\end{itemize}

\subsection{Internet of Things (IoT) Applications and their Characteristics}\label{IoT}

Internet of Things (IoT) refers to connected objects that collect and share data with other devices over the internet \cite{lin2017survey, jarwar2017exploiting}. It transforms conventional physical objects into smart objects. With the software/hardware improvements of the connected devices along with bandwidth and latency improvements in telecommunication technologies such as 5G/6G and other wireless technologies like Wifi6, the number and variety of connected devices are increasing rapidly \cite{chettri2019comprehensive, guo2021enabling, chen2020contention}. Cisco predicts that 500 billion devices will be connected to the internet by 2030, which would result in the generation of a massive amount of data that can be used to generate insights in a variety of IoT application domains such as smart healthcare, smart city, smart home, Industrial IoT (IIoT), smart agriculture and smart transportation \cite{zikria2021next, iyeswariya2020investigation}. 

Due to rapid growth in IoT data generation, applications have to undergo fast design, development and deployment cycles to both improve existing IoT services (i.e., moving to new tech stacks, AI augmentation and updating ML models, integration of novel data types, improvements in data processing method/algorithms, etc.) and introduce new services to the end-users. The ultimate vision of IoT is to move away from application development as "point solutions" and evolve towards an "IoT ecosystem" where cross-platform integration of heterogeneous devices, data, communication technologies, and application services occur \cite{woodhead2018digital, delicato2013towards}. To this end, IoT applications should be flexible to change, support service re-usability, and expose open interfaces to enable third-party integration of application services (i.e., big data analytic services, payment gateways, authentication and authorisation services, data visualisation services etc.) \cite{razzaq2020systematic}.

Due to improvements in Radio Access Network (RAN) technologies and consequent growth in mobile IoT devices (i.e., wearable gadgets, smartphones, vehicles, etc.), mobile IoT workloads keep increasing rapidly. To accommodate such fluctuating workloads, IoT application design has to mainly focus on the scalability of the application deployment \cite{venkatesh2017scalable}. Moreover, IoT applications consist of a large number of diverse services in terms of suitability of development technology stacks (i.e., I/O intensive services using NodeJs, machine learning using Python, etc.) and QoS expectations (i.e., bandwidth-hungry services like Multi-media Internet of Things (M-IoT), latency-sensitive applications like healthcare, tactile internet, etc.).

Due to these characteristics and requirements of IoT applications, their development is moving away from Monolithic architecture towards distributed application architectures, whereas deployment is moving away from cloud-centric deployment towards distributed computing paradigms like Fog computing.

\subsection{Edge and Fog Computing, Osmotic Computing and Related Paradigms}\label{fog}

\begin{figure}[h]
    \includegraphics[width=\linewidth]{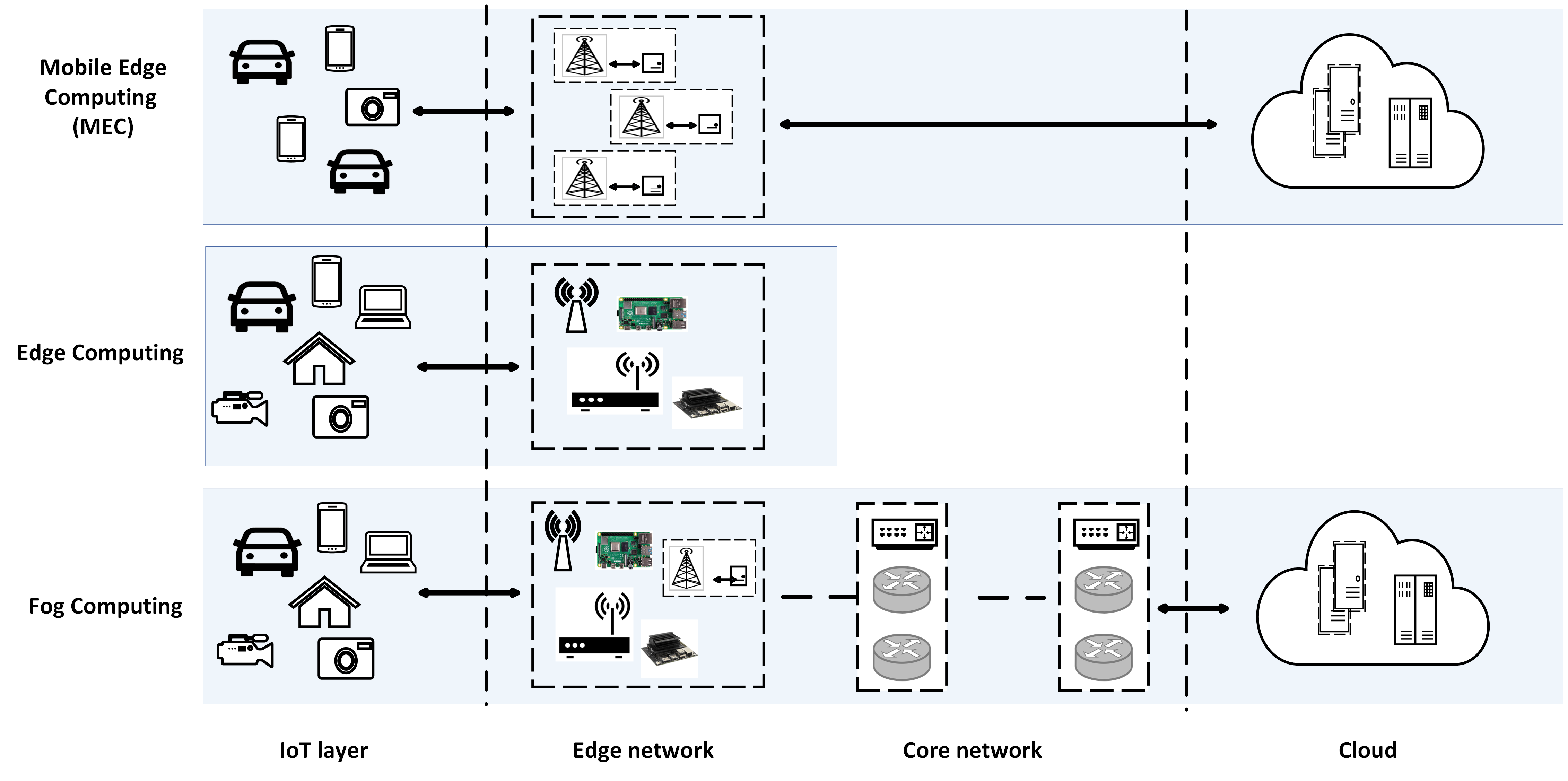}
    \caption{Fog computing, Edge computing and Mobile Edge Computing (MEC) }
    \label{fig:fogarchitecture}
    \vspace*{-3.5mm}
\end{figure}

Fog computing is a novel network computing paradigm which extends cloud-like services towards the edge of the network \cite{mahmud2018fog}. The fog computing paradigm was first introduced by Cisco in 2012 as a platform to support the unique requirements of IoT, such as low latency, location awareness, mobility support, and geo-distribution \cite{bonomi2012fog}. To this end, Fog computing introduces an intermediate layer between IoT devices and Cloud data centres \cite{mahmud2018fog}, which is organised in a multi-tier architecture. It exploits computation, storage and networking resources that reside within the path connecting end devices to the Cloud data centres \cite{ren2019survey}. Thus, Fog resources consist of a diverse set of resources (i.e., smart routers and switches, personal computers, edge servers, Raspberry Pi devices, micro-datacentres, cloudlets, etc.) that are distributed and resource-constrained compared to Cloud data centre resources. To overcome the resource limitations, the Fog computing paradigm maintains federated Fog computing architectures, where distributed Fog resources collaborate to satisfy client requirements \cite{alqahtani2021energy}. Moreover, the Fog computing layer maintains a seamless connection with the Cloud \cite{yousefpour2019all} so that computation-intensive tasks can be carried out using Cloud resources.

Edge computing and Mobile Edge Computing (MEC) computing also follow a similar concept to Fog computing, where their main objective is to move computation toward the users \cite{mahmud2020application}. Works such as \cite{mahmud2020application, yousefpour2019all, mahmud2018fog} define Edge computing as a paradigm limited to the edge network, which consists of devices such as mobile phones, access points etc. at the immediate first hop from IoT devices. In contrast, Fog extends this concept further to include the core network along with the integration of Cloud data centres when necessary to provide IaaS, PaaS and SaaS services in the proximity of the data sources. Mobile Edge Computing (MEC) focuses on mobile end-users and moves processing and storage capabilities to the edge of the mobile network by improving the capabilities of the base stations that reside within the Radio Access Network (RAN) of the 5G/6G networks \cite{dolui2017comparison}. However, all these paradigms share common characteristics, such as heterogeneity of resources, and limited and distributed resource availability, along with main goals such as lower latency, lower network congestion and mobility support.

Osmotic computing is a relatively new paradigm, first introduced in \cite{villari2016osmotic}, especially focusing on microservices-based application placement within Fog and Cloud environments. Osmotic computing aims to achieve a balanced deployment of microservices by utilising Edge/Fog and Cloud resources to satisfy the requirements defined at different levels of the microservices-based IoT application (i.e., service-level QoS requirements, microservice-level resource requirements etc.). Osmotic computing identifies MSA as a suitable approach for developing IoT applications for deployment within integrated Edge/Fog and Cloud environments due to their fine granularity, fast deployment and elasticity. Thus, this paradigm tries to address microservices-specific challenges such as microservice orchestration, networking, monitoring, elasticity control, and dynamic placement \cite{villari2016osmotic, androvcec2019systematic}.

\subsection{Microservices-based IoT Applications Placement vs DAG Workflow Scheduling}

\begin{figure}[h]
    \includegraphics[width=0.9\linewidth]{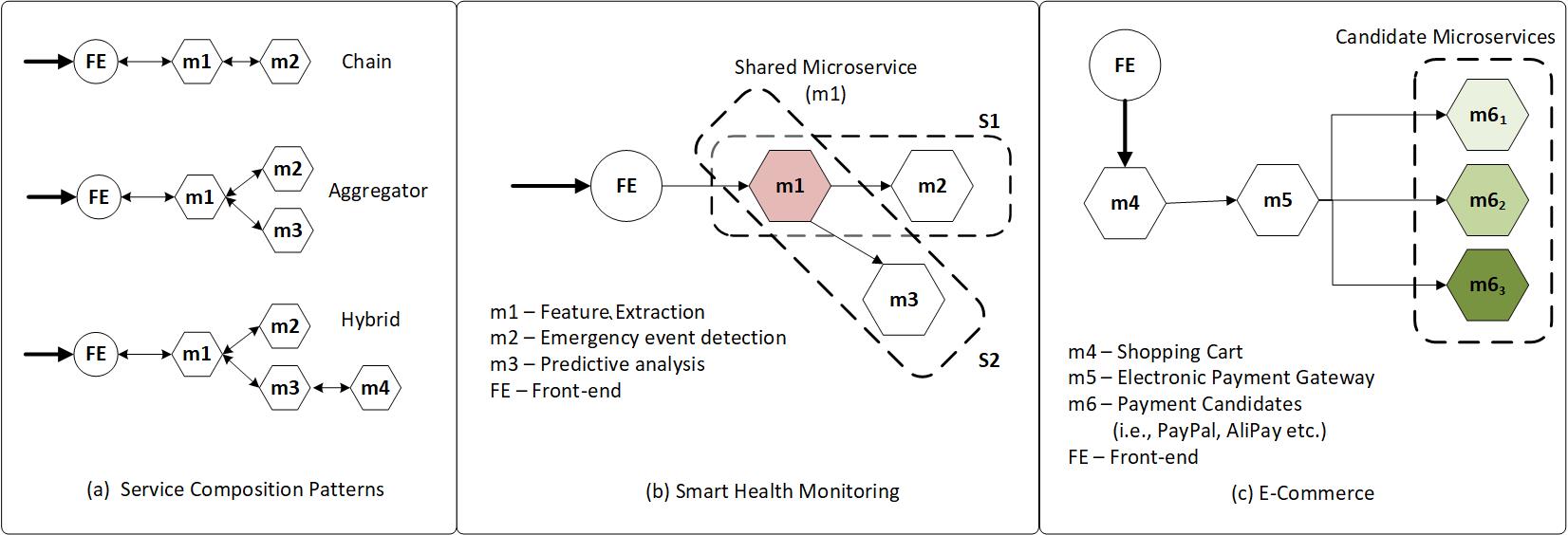}
    \caption{Characteristics of Microservices-based Composite Services}
    \label{fig:compositeServices}
    \vspace*{-3.5mm}
\end{figure}

In this section, we compare the "Microservices-based IoT application placement" problem with the Directed Acyclic Graph (DAG) workflow scheduling problem \cite{rodriguez2017taxonomy}, which is extensively addressed in the existing literature, considering both Fog and Cloud environments. We identify similarities and differences between the two to highlight the unique aspects related to the Microservices-based IoT application placement problem.

DAG-based workflows are used to model business processes as collections of distributed tasks where tasks, their dependencies and data flows are represented using DAGs \cite{rodriguez2017taxonomy}. Each DAG is characterised by a tuple, $\{V, E\}$, where $V$ and $E$ denote vertices and edges, respectively. In DAG-based workflows, vertices represent the tasks or sub-processes of the main process, and directional edges depict the data flow from the source task to the destination task, where the output from the prior task is fed as input for the next task. Thus, each DAG would consist of one or more starting tasks with no incoming edges and an ending task which provides the processed output. 

As microservice applications consist of interconnected microservices collaborating to perform domain logic, existing works model such applications as DAGs \cite{guerrero2019evaluation, faticanti2020deployment}, where microservices are represented by the vertices ($V$) of the DAG, and edges ($E$) represent the dependencies among microservices with direction from client microservices to invoked service. Thus, the directional edges denote microservice invocations. However, unlike in the case of DAG workflows, dataflows in microservice applications can not always follow a DAG due to the complex data dependencies among microservices, which can create cyclic dataflows. This can be further explained through the concept of composite end-user services in MSA. In IoT applications, "services" can be defined as functionalities accessed by the end-user expecting a certain output, and each application can consist of one or more such services \cite{pallewatta2022qos}. Fine-granularity of MSA creates composite services with different dataflow patterns such as chained, aggregator, and hybrid \cite{rathore2020investigations} (Fig.\ref{fig:compositeServices}a). Such patterns can result in cyclic data flows where one microservice invokes other microservices, receives responses from them and finally aggregates and/or further processes the responses before returning the results of the composite service.

Moreover, each IoT application consists of multiple composite services with heterogeneous QoS requirements. As an example, in a smart healthcare application for patient monitoring, there can be multiple services, such as emergency notification services that detect abnormalities in measured vitals in real-time (S1 in Fig. \ref{fig:compositeServices}b) and latency-tolerant services for predictive analysis to give early warnings about possible health issues to prompt preventive actions (S2 in Fig. \ref{fig:compositeServices}b) \cite{pallewatta2022qos}. Herein, large-scale IoT application placement has to consider competing QoS requirements of its services. These services can share some microservices among them, which further complicates the placements decisions (in Fig. \ref{fig:compositeServices}b Feature Extraction microservice - m1 is part of both S1 and S2). In contrast to this, workflow scheduling either defines QoS per workflow or an ensemble of workflows.

Each large-scale IoT application has to support a large user base of distributed users that expect ubiquitous access to the application. Thus, application placement includes challenges related to the application's scalability based on the number of users, their access locations and time. Moreover, the life cycle of the placement is also perpetual. In contrast to this, workflow scheduling is defined from a user perspective, where each user makes the placement request for their particular use and workflow deployment exists until the request from that client is processed, which makes the life cycle ephemeral. Moreover, MSA creates complex interaction patterns such as shared microservices among composite services, candidature microservices resulting in conditional dataflows based on the users, and third-party microservice usage.  \cite{zhao2020distributed} discusses electronic payment microservices where users can select one out of many payment options (i.e., PayPal, AliPay etc.) and the requests are directed to the microservice relevant to the selected option (Fig.\ref{fig:compositeServices}c). This results in multiple possible data paths within a single service and different levels of demand for each payment microservice based on the composition of the users accessing the service. Existing works on workflow scheduling rarely consider such interactions. 

Even though microservice applications can be modelled as DAGs, the above-highlighted characteristics differentiate microservices-based application placement problem from existing research focusing on DAG-based workflow scheduling within Fog environments. 

\begin{table}[t!]
	\caption{Comparison between Microservices-based application scheduling vs Directed Acyclic Graph (DAG) workflow scheduling}
	\label{table:Comparison2}
	\centering
	\footnotesize
	\resizebox{\linewidth}{!}
	{\begin{tabular}{p{0.5\textwidth}p{0.5\textwidth}}
			\hline
			\textbf{Microservices-based IoT applications Placement} & \textbf{DAG workflow Scheduling } \\
			\hline
		    Microservice invocation can be represented by a DAG  & Invocation and data flow are represented by a DAG \\
		    Multiple end-user services per app & Each DAG represents a single service \\
		     Microservices shared among services/ applications& No shared tasks among workflows \\
		    Per-service QoS with multiple services per app & QoS per workflow/ensemble of workflows \\
		    Life cycle : perpetual & Life cycle : mostly ephemeral \\
		    Service access shared among a large number of users & Mostly focus on a particular user \\
		    Complex interactions (shared microservices, candidate microservices, thrid-party microservices) & Not a main focus \\
		   \hline
	\end{tabular}}
	\vspace*{-5.5mm}
\end{table}
\end{document}